\documentclass[longnamesfirst,numberedappendix,apj]{emulateapj}
\usepackage[it,normalsize]{subfigure}
\usepackage{xspace}
\usepackage{epsfig}
\usepackage{amsmath}
\usepackage{apjfonts}
\usepackage{xr}
\renewcommand\arcsec{\mbox{$^{\prime\prime}$}\xspace}%
\renewcommand\micron{\mbox{$\mu$m}\xspace}%
\newcommand{\ssim}{\sim\!}

\newcommand{\kms}{${\rm km\;s}^{-1}$\xspace}
\newcommand{\msun}{$M_\sun$\xspace}

\newcommand{\msunyr}{$M_\sun$~yr$^{-1}$\xspace}
\newcommand{\secp}[1]{(\S\ref{#1})\xspace}
\newcommand{\sect}[1]{\S\ref{#1}\xspace}
\newcommand{\eqt}[1]{(\ref{#1})\xspace}

\newcommand{\panp}[1]{({\em #1})\xspace}
\newcommand{\pant}[1]{{\em #1}\xspace}
\newcommand{\pa}{P.A.\xspace}

\newcommand{\eg}{e.g.,\xspace}
\newcommand{\ie}{i.e.,\xspace}
\newcommand{\one}{II\xspace}
\newcommand{\two}{II\xspace}

\begin{document}

%%%%%%%%%%%%%%%%%%%%%%%%%%%%%%%%%%%%%%%%%%%%%%%%%%%%%%%%%%%%%%%%%%
%              Required TITLE Info                               %
%%%%%%%%%%%%%%%%%%%%%%%%%%%%%%%%%%%%%%%%%%%%%%%%%%%%%%%%%%%%%%%%%%

\shorttitle{A New View of SN~1987A }
\shortauthors{Sugerman {\em et al.}}

\title{A New View of the Circumstellar Environment of SN 1987A
 } 

\author{
 Ben E.\ K.\ Sugerman\altaffilmark{1,2},
 Arlin P.\ S.\ Crotts\altaffilmark{2,3},
 William E.\ Kunkel\altaffilmark{4},
 Stephen R.\ Heathcote\altaffilmark{5},
 and Stephen S.\ Lawrence\altaffilmark{6}
}

\altaffiltext{1}{Space Telescope Science Institute, 3700 San Martin
  Drive, Baltimore, MD 21218; sugerman@stsci.edu}
\altaffiltext{2}{Department of Astronomy, Columbia University,
  New York, NY 10027; arlin@astro.columbia.edu}
\altaffiltext{3}{Guest Observer, Cerro-Tololo Inter-American
  Observatory}
\altaffiltext{4}{Las Campanas Observatory, Carnegie Observatories,
  Casilla 601, La Serena, Chile; kunkel@ociw.edu}
\altaffiltext{5}{Southern Observatory for Astronomical Research, Casilla 603,
  La Serena, Chile; sheathcote@noao.edu}
\altaffiltext{6}{Department of Physics, Hofstra
  University, Hempstead, NY 11549; Stephen.Lawrence@hofstra.edu}

% Abstract
\begin{abstract}
We summarize the analysis of a uniform set of both previously-known
and newly-discovered scattered-light echoes, detected within 30\arcsec
of SN~1987A in ten years of optical imaging, and with which we have
constructed the most complete three-dimensional model of the
progenitor's circumstellar environment.  Surrounding the SN is a
richly-structured bipolar nebula.  An outer, double-lobed ``peanut,''
which we believe is the contact discontinuity between the red
supergiant and main sequence winds, is a prolate shell extending 28 ly
along the poles and 11 ly near the equator.  Napoleon's Hat,
previously believed to be an independent structure, is the waist of
this peanut, which is pinched to a radius of 6 ly.  Interior, the
innermost circumstellar material lies along a cylindrical hourglass, 1
ly in radius and 4 ly long, which connects to the peanut by a thick
equatorial disk.  The nebulae are inclined 41\degr south and 8\degr
east of the line of sight, slightly elliptical in cross section, and
marginally offset west of the SN.  The 3-D geometry of the three
circumstellar rings is studied, suggesting the northern and southern
rings are located 1.3 and 1.0 ly from the SN, while the equatorial
ring is elliptical ($b/a\lesssim0.98$), and spatially offset in the
same direction as the hourglass.  Dust-scattering models of the
observed echo fluxes suggest that between the hourglass and bipolar
lobes: the gas density drops from 1--3 cm$^{-3}$ to $\gtrsim0.03$
cm$^{-3}$; the maximum dust-grain size increases from $\sim0.2$\micron
to 2\micron; and the silicate:carbonaceous dust ratio decreases.  The
nebulae have a total mass of $\sim1.7$\msun, yielding a red-supergiant
mass loss around $5\times10^{-6}$ \msunyr.  We compare these results
to current formation models, and find that no model has successfully
reproduced this system.  However, our results suggest a heuristic
evolutionary sequence in which the progenitor evolves through two
``blue-loops,'' perhaps accompanied by a close binary companion.
\end{abstract}


%%%%%%%%%%%%%%%%%%%%%%%%%%%%%%%%%%%%%%%%%%%%%%%%%%%%%%%%%%%%%%%%%%
%                      introduction                              %
%%%%%%%%%%%%%%%%%%%%%%%%%%%%%%%%%%%%%%%%%%%%%%%%%%%%%%%%%%%%%%%%%%

\section{INTRODUCTION }\label{sec-intro}

Supernova 1987A, the first naked-eye supernova (SN) in four
centuries, was first discovered on 1987 Feb.\ 24.23 at the Las
Campanas Observatory (Shelton 1987), and was later identified with the
Large Magellanic Cloud (LMC) progenitor, Sk $-69$\degr202 \citep{San69},
a B3 I supergiant \citep{Rou78}.  From its spectral type and distance,
we infer a luminosity of $(3-6) \times 10^{38} \;{\rm erg}\;{\rm
s}^{-1}$, a surface temperature 15000--18000~K, and a radius $(2-4)
\times 10^{12}\;{\rm cm}$, which imply that the star exploded as a
blue supergiant \citep[or BSG;][]{Woo87}. Spectra taken shortly after
the explosion was discovered revealed Balmer lines, indicating a Type
II SN, but expanding at nearly $0.1c$.  A number of anomalies, such as
this high expansion velocity, a weak initial outburst, a very hot UV
spectrum at early times, and subsequent rapid color evolution,
indicated a compact progenitor envelope, typical of a BSG, rather than
the expected red supergiant (RSG) progenitor.  Heavy mass loss could
have driven Sk $-69$\degr202 to the BSG stage (mass loss being reasonable
given the large number of Wolf-Rayet stars in the LMC), although
evidence for a substantial envelope (the long slow rise of the optical
light source, the late appearance of near-IR H lines and a lack of
early escape X and $\gamma$-rays) also exists.  For a  complete
accounting, see the review by \citet{Arn89}.

\citet{Pan87} reported narrow UV emission lines in {\em International
Ultraviolet Explorer} spectra taken on day 80, which they interpreted
as arising from pre-existing CS material.  An analysis of these
spectra by \citet{Fra89} revealed a large overabundance of nitrogen,
suggesting both significant CNO processing by the progenitor, and that
a reasonable fraction of the progenitor's hydrogen envelope had to
have been ejected to reveal such material.  This argues that the star
first evolved into a RSG before returning to the blue in a so-called
``blue loop'' \citep{Hof64}.  Both RSG \citep{HD78} and BSG
\citep{HM84} stars are known within the LMC, thus this evolutionary
sequence is consistent with known populations.

\begin{figure}\centering
\includegraphics[width=2.in,angle=0]{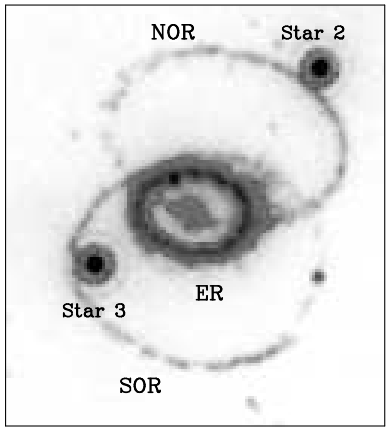}
\caption{{\em HST} WFPC2 negative image of a $4\farcs5\times5\farcs0$
field surrounding SN~1987A, taken in F656N (H$\alpha$ and
\ion{N}{2}). North is up, east is left.  The ER (central ring)
surrounds the ejecta from the SN (center), and is flanked by the north
and south ORs (NOR and SOR, respectively).  The companion stars 2 and
3 \citep{WS90} are positionally coincident along the line of sight.
To increase the display range, the ER has a separate
color stretch.
\label{PCMAP}}
\end{figure}

The first resolved images of this CS material were taken on days 750
\citep{CKM89} and 1027 \citep{Wam90a}, showing a central, elliptical
structure, surrounded by two outer loops \citep{CKM89,Wam90b}.  The
geometry of this material was the subject of some debate, remaining
poorly understood until {\em Hubble Space Telescope} ({\em HST})
imaging \citep[][and references therein]{Bur95} clearly revealed a
dense circumstellar equatorial ring (ER) flanked by two larger outer
rings (ORs), as shown in Figure \ref{PCMAP}.  To explain how a
mass-losing supergiant produced this nebula, most authors invoke the
interacting stellar winds scenario \citep{Kwo82,Bal87}, in which
colliding winds from the BSG and RSG mass-loss stages conspire to
create three ring-like overdensities within the CS environment (CSE).
These were ionized by the SN light pulse and are observed today
through recombination cooling.

Since the SN lightcurve was strongly peaked for a finite duration, and
the progenitor is surrounded by a structured CSE, we expect that the
light pulse will scatter off of surrounding reflective material (\eg
dust), producing ``light echoes'' that are observable long after the
SN has faded \citep{Cou39}.  \citet{Bon89}, \citet{CM89}, and
\citet{CK89} discovered a light echo at $\ssim 9\arcsec$ whose
position agrees well with the predicted location of a contact
discontinuity (CD) between the RSG outflows and the interstellar
bubble formed by the early-type blue MS progenitor's winds
\citep{CE89}.  Crotts et al.\ \citep[\eg][hereafter CKH95]{CK91,CKH95}
have continued to search for light echoes in a regular campaign using
emission-suppressed continuum imaging.  They have found echoes
indicative of the structure of the three-ring nebulosity, forming an
equatorial waist and sides of an hourglass, yet ending abruptly near
the ORs with no capping surface.  Outside this inner nebula, they
found four additional features: (1) a reconfirmation of the CD light
echo of \citet{Bon89} at $9-15\arcsec$; (2) a sheet along the
equatorial plane of the ER, bisecting the nebula; (3) diffuse flux
from the hourglass nebula to the CD; and (4) a discontinuity in this
nebula, called ``Napolean's Hat'' \citep{Wam90a}.

Observation and modeling of this circumstellar environment provide a
unique opportunity to reconstruct the mass-loss history of the
progenitor.  The SN~1987A nebula is one example of a much larger class
of bipolar outflows, which occur in nearly all evolved stars
(\eg luminous blue variables, B[e] stars, AGB mass loss, bipolar and
asymmetric planetary nebulae).  This
``stellar paleontology'' is of immense importance for
understanding bipolar-outflow and bipolar-nebula formation
mechanisms, not just of this object or in intermediate-to-high mass
stars, but in stars of all masses.

We summarize in this paper an observational effort, at the highest
sensitivity and resolution to date, to understand more fully the
entire circumstellar environment of SN~1987A, and to recreate a more
complete history of the progenitor's mass-loss, as revealed by light
echoes in 16 years of optical imaging.  After introducing echoes in
\sect{sec-le}, we summarize the data and reduction in
\sect{sec-echodata}.  These data are used to build a complete model of
all material within $\sim$30 ly of the SN, the 3-D analyses of which
are summarized in \sect{sec-CSE}, followed by the dust properties in
\sect{sec-density}.  Full details of these sections are presented in
\citet[][hereafter Paper \one]{Sug05}.

With a much greater volume of the progenitor's CSE revealed, we find
that no extant model adequately explains how that environment was
formed.  In \sect{sec-echointerp}, we review and critique formation
scenarios for this system, including the progenitor's MS wind,
interacting supergiant winds, wind-compressed disks, and the influence
of a binary companion.  We use the light-echo data to constrain the
parameter space of these models, and we qualitatively sketch a hybrid
formation sequence for the entire CSE in \sect{sec-evol}.
 It is our hope that the results presented here will
serve as a significantly-improved set of constraints for hydrodynamic
models of the CSE formation, for stellar-evolution models of the
progenitor, and for addressing the general question of asymmetric and
bipolar stellar outflows.

%%%%%%%%%%%%%%%%%%%%%%%%%%%%%%%%%%%%%%%%%%%%%%%%%%%%%%%%%%%%%%%%%%
%  le.tex                                                        %
%%%%%%%%%%%%%%%%%%%%%%%%%%%%%%%%%%%%%%%%%%%%%%%%%%%%%%%%%%%%%%%%%%

\section{Light Echoes }\label{sec-le}

\begin{figure}\centering
\includegraphics[angle=0,width=3.25in]{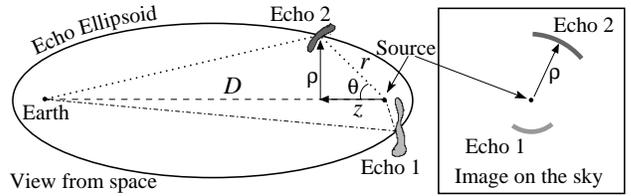} 
\caption{Cartoon schematic of a scattered-light echo.  Echoes appear
as arcs or rings on the plane of the sky, with a one-to-one mapping
between their 2-D image positions (right) and their 3-D locations in
space (left).  N.B.\ Distances are not to scale, since $D\gg z$.
\label{echo_toon}}
\end{figure}

When a light pulse is scattered by dust into the line of sight, an
observable echo is produced, provided the pulse is sufficiently
luminous and the dust sufficiently dense.  An echo observed a given
time after the pulse must lie on the locus of points equidistant in
total light travel from the source and observer, that is, an ellipsoid
with known foci.  This simple geometry, shown in Figure
\ref{echo_toon}, directly yields the line-of-sight depth $z$ (and
hence the three-dimensional, or 3-D, position)
of an echo, uncertain only by the assumed distance $D$ to the source.
For $D\gg r$, one finds \citep{Cou39}
\begin{equation} \label{sec-le-sb1}
 z=\frac{\rho^2}{2ct}-\frac{ct}{2}
\end{equation}
where $\rho=r\sin{\theta}$ is the distance of the echo from the
source in the plane of the sky, $\theta$ is the scattering
angle, and $t$ is the delay between observing the echo and unscattered
light pulse. 

A complete discussion of light echoes, including single-scattering
models and the observability of echoes around a wide variety of
cataclysmic and other variable stars, can be found in \citet{Sug03}.
The relevant light echo model, dust and outburst properties for SN
1987A are summarized in \sect{I-sec-le} of Paper \one.  In general,
echo brightness scales directly with dust density, inversely with
distance, and decreases non-linearly with increasing scattering
angle. 

%%%%%%%%%%%%%%%%%%%%%%%%%%%%%%%%%%%%%%%%%%%%%%%%%%%%%%%%%%%%%%%%%%
% echodata.tex
%%%%%%%%%%%%%%%%%%%%%%%%%%%%%%%%%%%%%%%%%%%%%%%%%%%%%%%%%%%%%%%%%%

\section{Data and Reduction} \label{sec-echodata}

\subsection{Observations \label{sec-echodata-obs}}

SN~1987A has been observed many times per year since day 375 (after
core collapse) in a ground-based campaign to monitor the appearance
and evolution of its light pulse as it illuminates circumstellar and
interstellar material.  To minimize confusion from sources of
nebular-line emission, four specially-selected continuum bands were
used for the monitoring campaign, centered at 4700, 6120, 6880, and
8090\AA.  Some data from this campaign have previously been reported
by \citet{CKM89}; \citet{CH91}; \citet{CK91}; and CKH95 in studying
the circumstellar environment, and in \citet{Xu94,Xu95} and
\citet{Xu99} for interstellar echoes.  Seventeen epochs of imaging
with high signal-to-noise, good seeing (arcsec or better), and
sufficient resolution to resolve most crowded stars exist between 1998
Dec and 1996 Jan.  These data are supplemented with archival {\em HST}
Wide Field and Planetary Camera 2 (WFPC2) imaging taken between 1994
and 2001.  Full details of the observations are given in
\sect{I-sec-echodata} of Paper \one.

\subsection{Data Reduction }\label{sec-reduc}

\begin{figure*}\centering
%\begin{figure}\centering
\includegraphics[width=5.5in]{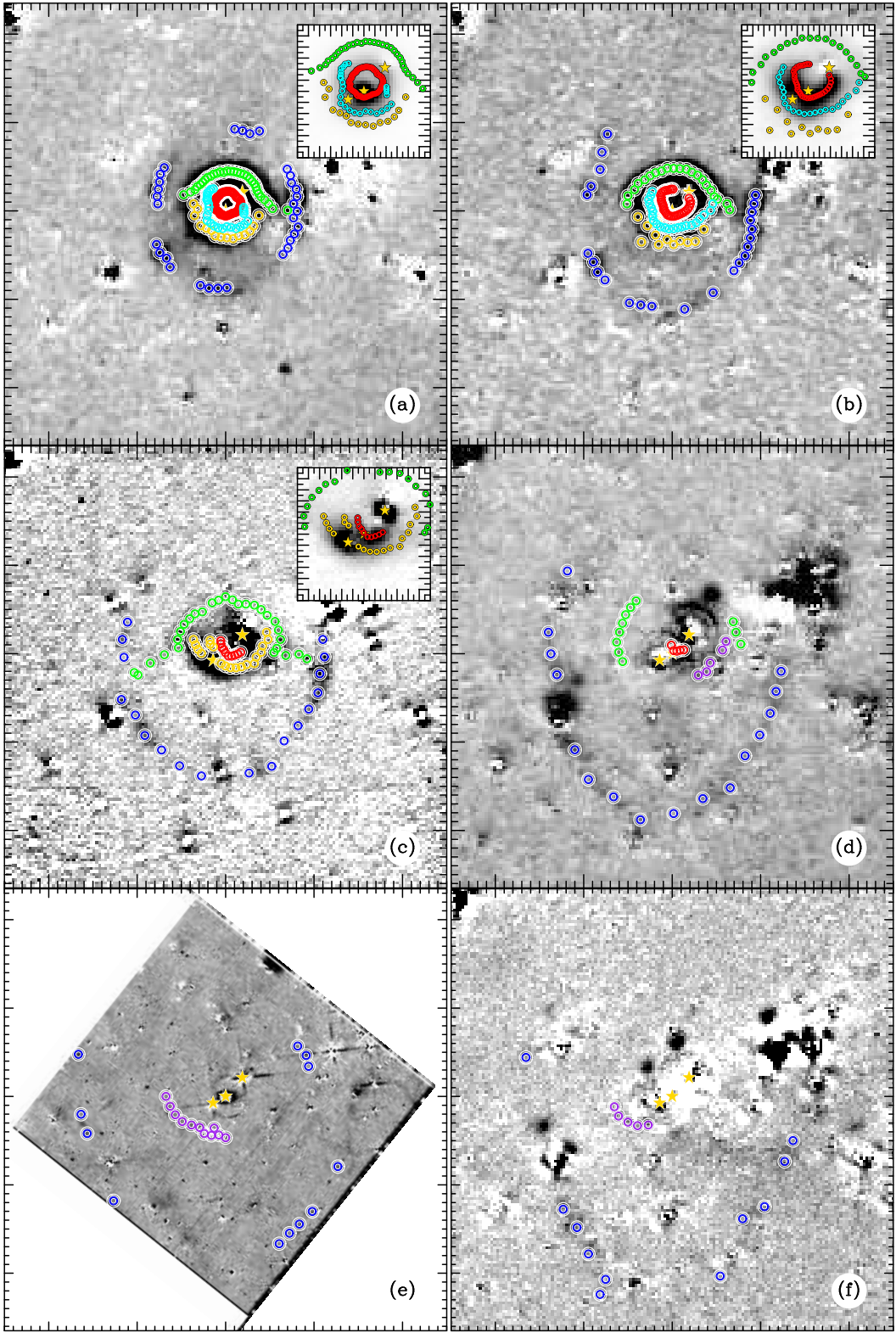}
\caption{Selected $50\arcsec\times50\arcsec$ difference images.  North
is up, east is left, and major ticks mark 10\arcsec.  The position of
the SN, Stars 2 and 3 are marked with yellow stars.  The inset shows
the central 12\arcsec at a different color stretch to resolve the
innermost echoes.  Echoes are marked by colored circles, explained in
\sect{sec-lea}.
\panp{a} 6067\AA\ image from day 750.
\panp{b} 6120\AA\ image from day 1028.
\panp{c} 6120\AA\ image from day 1469.
\panp{d} 612\AA\ image from day 2095.
\panp{e} WFPC2 image from day 2769.
\panp{f} 612\AA\ image from day 2874.
\label{dimages}}
%\end{figure}
\end{figure*}

The complete data-reduction is presented in \sect{I-sec-reduc} of
Paper \one.  To search for light echo signal, we employ a customized
implementation of the PSF-matched difference imaging \citep{TC96} IRAF
package {\em difimphot}, fully described in \sect{I-sec-dip} and
Appendix A of Paper \one.  A subset of the PSF-matched difference
images are shown in Figure \ref{dimages}; see \sect{I-sec-lea-data} of
Paper \one for additional data and discussion.  The innermost echoes were
particularly bright up to day 1469, and are washed out at the
brightness-stretch needed to display the outer echoes.  In such cases,
a 12\arcsec inset centered on the SN is shown at the top right of each
panel, scaled in brightness to better-resolve the innermost echo
signal.

%%%%%%%%%%%%%%%%%%%%%%%%%%%%%%%%%%%%%%%%%%%%%%%%%%%%%%%%%%%%%%%%%%
% lea.tex
%%%%%%%%%%%%%%%%%%%%%%%%%%%%%%%%%%%%%%%%%%%%%%%%%%%%%%%%%%%%%%%%%%

\subsection{Echo Measurement and Visualization }\label{sec-lea}

Echoes were detected and measured in difference images by fitting a
series of convolved moffats to radial surface-brightness
profiles of arclength 10\degr centered on the SN.  See
\sect{I-sec-lea-data-echoes} of Paper \one for complete details.  
The centers of all moffats have been marked in the difference
images (Fig.\ \ref{dimages}), and color coded to compare echoes
believed to be physically associated.

During their detection, echoes were categorized as belonging to one of
the three expected structures: (1) a circumstellar hourglass-shaped
nebula reported in CKH95, (2) Napoleon's Hat \citep{Wam90a}, or (3)
the discontinuity between the progenitor's RSG and MS winds
\citep{CE89}.  In Figure \ref{dimages}, these have been color coded
red, green, and blue, respectively.  However we also found 
additional echoes between the hourglass and
Napoleon's Hat, arbitrarily colored cyan for those looping to the north
of the SN, or gold to the south.  In later epochs, many of
these southern echoes have positions that are more consistent with
a counterpart to Napoleon's Hat, and have been marked 
purple.

That diffuse structure exists between the inner hourglass and
Napoleon's Hat has been previously reported by \citet{CK91} and
CKH95. These are consistent with the cyan and yellow echoes identified
in the same region of the current data, which now map out a much more
extensive volume of the innermost CSE.
Contact-discontinuity echoes have been previously reported in
observations only up to day 1028, while we are able to trace this
structure through day 3270.  Napoleon's Hat was imaged as early as day
850 and up to day 1650 by \citet{WW92}, while we detect its signal
from day 659 to 2095.  We believe this is the first discussion of the
southern-counterpart (purple) to Napoleon's Hat, which was illuminated
between days 1787--3270.

\subsection{Echo Visualization \label{sec-lea-analy}}

\begin{figure}
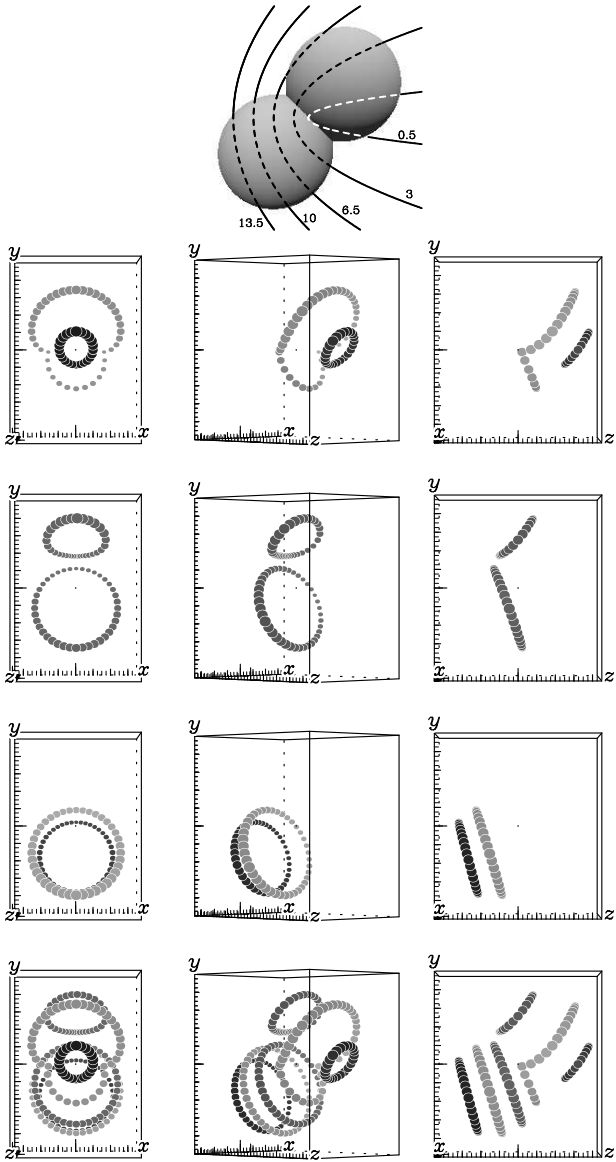
\centering
 \includegraphics[width=1.2in,angle=-90]{f04a} \\
 \includegraphics[width=3.25in,angle=0]{f04b}
 \caption{Examples of 3-D renderings of light echoes in a simple
 bipolar nebula, depicted at top, which is inclined 45\degr to the
 line of sight (to the right).  Each lobe is 5 ly in radius, with the
 centers separated by 8 ly.  Five echo parabolae are drawn, each
 occuring at the time (in years) indicated next to each curve.  Below,
 renderings show the light echoes for the first two parabolae (first
 row), third parabola (second row), last two parabolae (third row),
 and all parabolae (last row).  Axes are defined as $z$ toward the
 observer and $x-y$ in the plane of the sky, with $x$ increasing west
 and $y$ north.  Axis labels indicate the positive direction.  The
 origin is set at the SN, and is always marked by a black dot.  Major
 ticks mark 2 ly, and the origin is indicated by the longest tick
 along each axis.  The coordinates and axes have been given a slight
 perspective transformation (Appendix C of Paper \one).  Points are
 shaded using simple ray tracing, and larger points are closer to the
 observer.  {\em Left Column:} Face-on view (the plane of the
 sky). {\em Middle Column:} Oblique view rotated 45\degr. {\em Right
 Column:} Side view from far to the east.  Note that the geometry of
 the nebula, as revealed in light echoes, is only clear when many
 echoes well-separated in time are considered at once.
 \label{3de_parab}}
\end{figure}

To study echo positions in 3-D, we have custom-written graphics
software to perform simple renderings, in which the positions of the
echoes are transformed to allow viewing from any angle.   The method
is fully explained in \sect{I-sec-lea-analy-3de} of Paper \one, and
briefly annotated in the caption of Figure \ref{3de_parab}, which
shows an illustrative example.

The region of space probed by a light echo is an unusual geometric
function, and we are generally unfamiliar with viewing structures by
their intersection with parabloids.  To facilitate the understanding
of this mapping, we have rendered echoes from a bipolar nebula in
Figure \ref{3de_parab}. These are intended to guide the reader in
visualizing such intersections, and in translating from 2-D data on
the sky (left column) to 3-D positions in real space (right columns).
This also shows that the geometry of a structure can only be deduced
from multiple echoes.  See \sect{I-sec-lea-analy-parabs} of Paper \one
for more detailed discussion, and a map of the parabolae corresponding
to all epochs of data.

\begin{figure*}\centering
%\begin{figure}\centering
\includegraphics[angle=0,width=6 in]{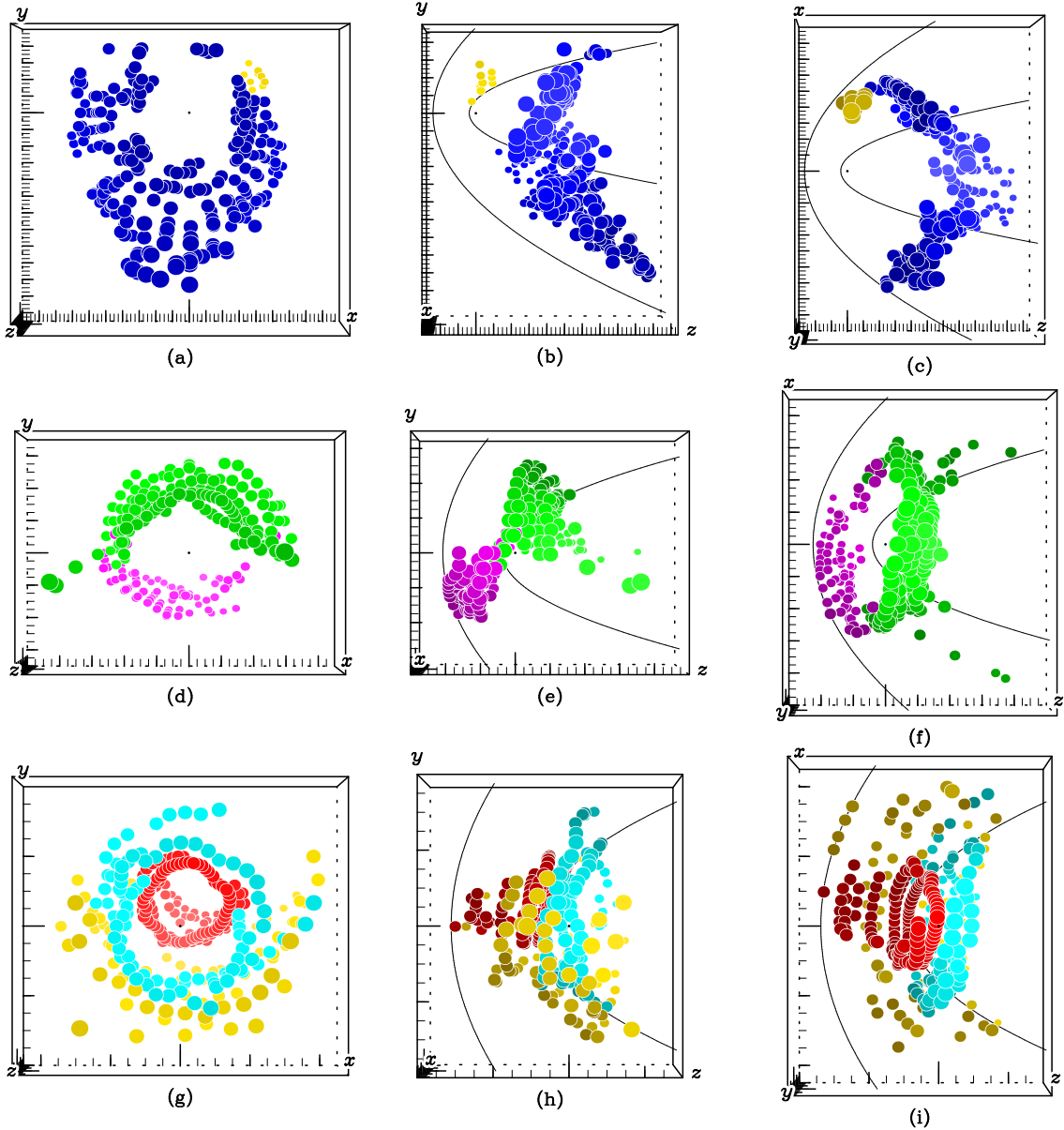}
\caption{Rendered views of all light echoes identified in Fig.\
 \ref{dimages}.  Points have been rendered using the method described
 in \sect{sec-lea-analy}; also see Fig.\ \ref{3de_parab}.  The
 left column shows observed views in the plane of the sky, the middle
 column shows views from the side (far to the east) and the right
 column shows views from the top (far to the north).  Parabolae from
 the earliest and latest epochs at which echoes were observed are
 indicated in the side and top views.  Point colors correspond to
 those in Fig.\ \ref{dimages} except for the yellow points in the top
 row, which denote echoes found in WFPC2 images which could not be
 resolved from the residuals of nearby bright stars in ground-based
 images.  Note that the field-of-view changes between rows, but major
 tick marks always denote 2 ly.  
 {\em Top row:} Contact discontinuity echoes.  
 {\em Middle row:} Napoleon's Hat and southern counterpart.
 {\em Bottom row:} Inner circumstellar hourglass echoes.  
 \label{3dle}}
%\end{figure}
\end{figure*}

Echo positions in 3-D are computed from their epoch and 2-D image
positions using equation \eqt{sec-le-sb1}, assuming the echo center is
produced by the light pulse maximum at day 87.  We adopt a distance
to the SN of 50 kpc, a common average between many of the derived LMC
distances \citep{Gou98,Fea99,Rom00}.

The 3-D positions of all echoes are rendered in Figure \ref{3dle}.
The top, middle, and bottom rows show the contact discontinuity (CD),
Napoleon's Hat (NH), and the inner circumstellar (CS) material,
respectively.  Colors correspond to those indicated in Figure
\ref{dimages}, except for the yellow points in the CD, which were
identified in WFPC2 images very close to a bright star cluster
northwest of the SN (\pa 300\degr, $\rho=15\arcsec$).  We distinguish
these data because large residuals from the closely-spaced, bright
stars made it impossible to verify the echoes in ground-based images.

\section{A Complete Picture of the Circumstellar Environment}\label{sec-CSE}

A rigorous analysis of the 3-D structure of the light echo data is
presented in Paper \two, \S\ref{I-sec-CD}--\ref{I-sec-CS}, however
we describe the results in the following subsections.  We begin
with the reasonable assumption that the echoes lie on continuous
surfaces, only subsets of which have been illuminated.  
Echo brightness decreases with increasing scattering angle and
distance from the illuminating source, and increases with dust number
and column density.  That echoes were not observed in some regions
between the parabolae in Figure \ref{3dle} need not imply lack of
scattering dust, but may result from unfavorable geometry, or limitations in the data quality.

%%%%%%%%%%%%%%%%%%%%%%%%%%%%%%%%%%%%%%%%%%%%%%%%%%%%%%%%%%%%%%%%%%
% sec-CD Contact Discontinuity
%%%%%%%%%%%%%%%%%%%%%%%%%%%%%%%%%%%%%%%%%%%%%%%%%%%%%%%%%%%%%%%%%%
\subsection{Contact Discontinuity Echoes }\label{sec-CD}

When examined in spherical-polar coordinates, the CD echoes (Fig.\
\ref{3dle}\pant{a--c}) appear to trace two different geometric shapes.
Points east and west of the SN form a shell-like feature at roughly
constant radius, while points south of the SN (\pa 150\degr--210\degr)
form a radial feature at roughly constant inclination.  In both cases,
the points appear to lie on a surface with some degree of symmetry,
such as a spheroidal or conical structure.  Since only a partial cross
section of the CD nebula has been illuminated, finding its inclination
through numeric minimization of a merit function (e.g.\ the
best-fit cone or line; see Appendix B of Paper \one) was not highly
successful.  We supplemented those results by testing how the
echo points compared to geometric figures observed at
different orientations, and conclude that the CD echoes lie roughly on
a surface of rotation about a single axis inclined 40\degr south and
8\degr east of the line of sight.  This  axis may also be offset
east of the SN by $\lesssim1$ ly.  The cross section gives some
indication of an ellipticity of $b/a=0.95$, with the major axis
aligned toward north, however this is not a robust measurement.

\begin{figure}\centering
\includegraphics[height=3.25in,angle=-90]{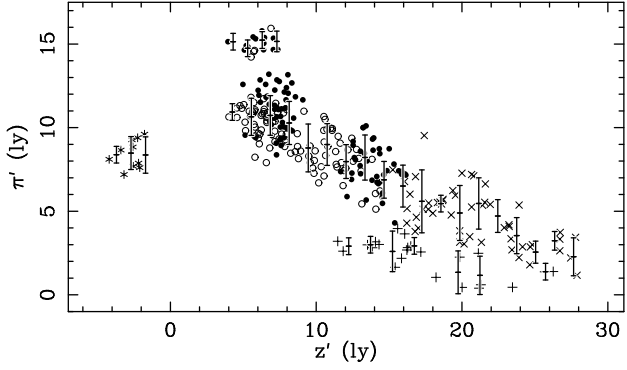}
\caption{CD echoes, transformed into cylindrical coordinates with $z'$
along the axis of symmetry (40\degr south, 8\degr east) and 
$\pi'$ measured radially from it.  Eastern and western points are
marked by filled and open circles, respectively.  Southern points are
marked as ``$\times$'' except for a subset at smaller $\pi'$ marked
with ``+''.  A small subset of WFPC2 echoes to the northwest (yellow
points in Fig.\ \ref{3dle}) are marked with asterisks.  Overplotted is
the average radial profile, measured as the average position (and
standard deviation) of points binned along $z'$.
\label{plCD4}}
\end{figure}

The data are transformed into a ``primed'' cylindrical coordinate
system about this axis, designated by $z'$ along the axis and $\pi'$
the radius.  Figure \ref{plCD4} shows the resulting radial profile
$\pi'(z')$.  Assuming these data trace a structure that (1) has
rotational symmetry about $z'$, and (2) is symmetric about the equator
($z'=0$), an average radial profile of $\langle\pi'\rangle$ versus
$z'$ can be constructed by binning the points along the axis.  This
(and standard deviations in each bin) is shown over the individual
data.  Some circles located at $\pi'\sim 15$ and $z$=4--8 ly appear
disconnected from the rest of the shell-like material.  These points,
which we call the ``spurs,'' were treated separately when making the
average radial profile.

The probable CD structure is visualized by revolving this function
about the symmetry axis, reflecting it about the equator, and
reinclining it to the favored orientation, as shown in Figure
\ref{3dpl}\pant{a--b}.  Colors are the same as Figure \ref{3dle}, only
the southern echoes have been colored red and green to distinguish
points marked with ``$\times$'' and ``+'' in Fig.\ \ref{plCD4}.
Figure \ref{3dpl}\pant{c--d} shows the complete structure in monotone
grey, overlaid with the actual echo points from Figure \ref{3dle}, to
show exactly which parts of the complete structure were sampled.

The complete structure is fairly complicated, and can not be described by a
single geometric function.  Figure \ref{toons}\pant{a} shows a scaled
cartoon of the salient features.  A ``shell'' lies along a prolate
spheroid with a polar axis of 20 ly and equatorial axes of 11 ly.
However, the ends of this spheroid have been drawn out into tapering,
``radial'' cones with opening angles of about 35\degr, extending from
16 to 28 ly from the SN.  Embedded within this prolate structure are
narrower, tapering cones (the ``jet'') extending from 10--26 ly from
the SN, with an opening angle of about 20\degr, and a maximum radius
of 3 ly.  The ``spurs'' lie along a cylindrical annulus that smoothly
encircles the CD.

It is unclear whether the prolate shell is continuous in $z'$ along
its equator (dashed equatorial lines in Fig.\ \ref{toons}\pant{a}),
since no echoes were observed from that region.  Whether this is due
to shadowing from material closer to the SN is addressed in
\sect{I-sec-density-density-shadowing} of Paper \two.  The
northwestern-WFPC2 echoes are positioned at $z'\sim4$ ly along the
axis, and at roughly 8 ly in radius.  This could suggest the prolate
shell is pinched at its waist to a smaller radius of 8 ly.  However
given the very limited spatial sampling of these inner echoes, there
is little evidence that they lie on a uniform structure.  Similarly,
very little signal was detected from the spurs which may suggest that
they are isolated clumps, rather than a uniform feature.

\subsection{Napoleon's Hat}\label{sec-NH}

Figures \ref{3dle}\pant{d}--\pant{f} show the 3-D positions of the NH
echoes.  The northern (green) points forming the
familiar bow-shape for which this structure was named \citep{Wam90a}
appear to lie on a thin, cylindrical shell that is completed by the
newly-discovered southern points.  To avoid confusion between
this and previous work, we will keep the name ``Napoleon's Hat,'' by
which we refer to the entire ensemble of echoes.  Despite the
temptation to call the southern points ``Napoleon's Collar,'' we will
simply refer to these as NH-north and NH-south, when needed.

Two nearly-horizontal features protrude from this shell toward the
observer, appearing horn-like when viewed from above (panel \pant{f}).
In the following discussion, we refer to these as the ``horns,'' to
distinguish them from the rest of the NH material.  These are the
echoes seen in early imaging that bridged between NH-north and the
larger-radii CD echoes (Fig.\ \ref{dimages}).  

Viewed in spherical-polar coordinates, the NH echoes are consistent
with a cylinder or hourglass.  We measured the shape and inclination
of such structures by fitting a biconical frustum to all points, as
explained in Appendix B.4 of Paper \one.  The best fit is an hourglass
with elliptical cross section, oriented  40\degr south and 7\degr
east of the line of sight.  The semi-major and minor axes are 4.6 and
3.8 ly, with opening angles at the axes of 28 and 40\degr,
respectively, and the major axis is rotated $92-101\degr$ east of
north.  That the cross-section is elliptical is unexpected, and we
reverified this by directly examining the data looking down the
axis of inclination.  The best-fit ellipse (Appendix B.2 of
Paper \one) to the resulting distribution has negligible offset from
the origin, and $b/a=0.82\pm0.02$ with the major axis rotated
$103\degr\pm 3$ east of north.

\begin{figure}\centering
\includegraphics[height=3.25in,angle=-90]{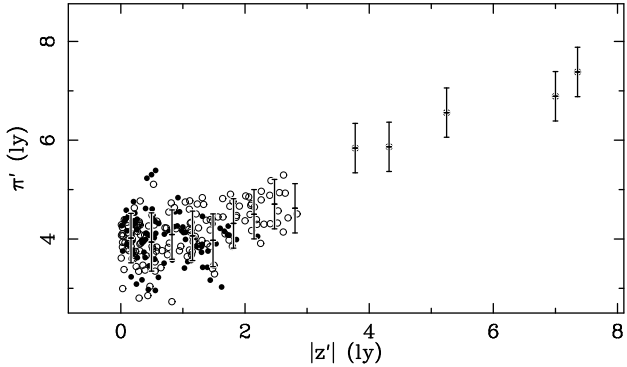}
\caption{As Fig.\ \ref{plCD4}, but for NH data, with the ellipticity
removed.  Overplotted is the average radial profile, measured in bins
along $|z'|$.  Single points have fixed errors. 
 \label{plNH4}}
\end{figure}

Removing this ellipticity from the radial values of $\pi'$ as a
function of $z'$ yields the distribution shown in Figure \ref{plNH4}.
Also shown is the average radial profile of the NH echoes, generated
by binning the points along the inclination axis and computing the
average radius about that value.  As with the CD, we visualize the
probable structure containing these echoes by reflecting this average
radial profile about the equator and revolving it around the axis. The
result is an inclined hourglass, as rendered in Figure
\ref{3dpl}\pant{e--h}.

An examination of Figure \ref{3dpl}\pant{h} shows that very little of
the hourglass was actually probed by the horns.  We show in
\sect{I-sec-NH-geom-horns} of Paper \two that such limited illumination
was a natural result of the scattering geometry of this complete
structure.  The hourglass therefore flares to a radius (or semi-minor
axis if the cross section remains elliptical) of $\sim 7$ ly about 7
ly from the SN along the axis of symmetry.  

The complete cross section is shown in the cartoon sketch in Figure
\ref{toons}\pant{b}.  Of note are the radial features in, and about 1.5 ly
above/below, the equatorial plane.  CKH95 reported a
small set of echoes within 3 ly of the SN that are coincident with the
ER plane, which they interpreted as evidence for an extended
circumstellar equatorial disk.  These NH equatorial echoes appear to
extend this disk to a radius of 5 ly, with a thickness of $\gtrsim
0.75$ ly.  We note that this structure is inconsistent with NH being a 
parabolic bow-shock north of the SN, as proposed by \citet{WW92} and
\citet{WDK93}.  This is discussed in greater detail in
\sect{I-sec-NH-comp} of Paper \two.

\subsection{Inner Circumstellar Material}\label{sec-CS}

\begin{figure}\centering
\includegraphics[width=3.25in,angle=0]{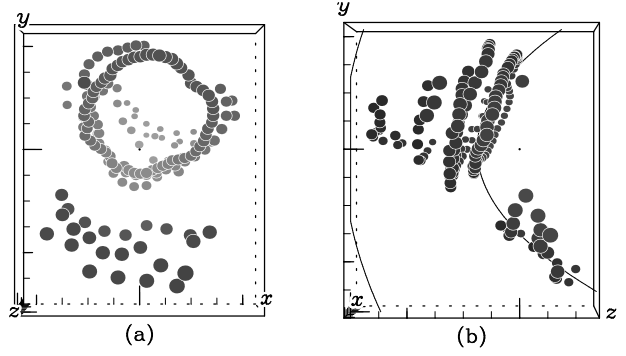}
\caption{CS points believed to lie along the inner hourglass.
 \label{3de_CS}}
\end{figure}

The inner 3\arcsec surrounding SN~1987A have been previously studied
by CKH95 using a subset of the data presented here, in which the
authors reported a double-lobed CS hourglass with a distinct
cylindrical symmetry axis.  Although the red echo points in Figure
\ref{3dle}\pant{g--i} appear to lie along such a surface, we cannot
immediately explore this feature in our data since it is unclear
which extended-flux echoes (cyan and gold points) are associated with
an hourglass.  We use the work of CKH95 as a springboard to help
disentangle our data.

Fitting a biconical hourglass to the CKH95 data, we find the structure
is inclined 45\degr south and 8\degr east of the line of
sight.  This compares favorably to the orientation they reported using
a minimization-of-scatter estimator.  Adopting this as a preliminary
orientation, we transformed our CS echo points into cylindrical
coordinates about this axis, which revealed an additional subset 
consistent with those initially identified as part of the
hourglass.  The full set of probable hourglass  points
is rendered in Figure \ref{3de_CS}.  As can be seen in panel
\panp{b}, only a small arc along the southernmost limb of of the
southern lobe was probed by the earliest observed echoes.

A combination of fitting biconical hourglasses to these data, and
ensuring that the inner waist of this hourglass reproduces the
observed shape of the ER, yields the 3-D geometry of both structures.
We find the hourglass is inclined 41\degr south and 8\degr east of the
line of sight, with its axis shifted $0.1$ ly west of the SN.
The waist has a semi-major axis of $1.04$ ly rotated
$9\degr$ north of east, with $b/a=0.94$, and the
frusta have half-opening angles of $12$.  Note that this is
consistent with the findings by \citet{Sug02} that the nebula is
offset slightly west of the SN.

Deprojected by the above inclination, the ER is best-fit by an ellipse
with major axis 0\farcs82 rotated 9\degr north of east, $b/a=0.98$,
and the centroid is shifted 19 mas west of the SN. This is the first
direct measurement of the ER's orientation and deprojected geometry.
Full details of these analyses are presented in \sect{I-sec-CS-geom} of
Paper \two.

\begin{figure}\centering
\includegraphics[height=3.25in,angle=-90]{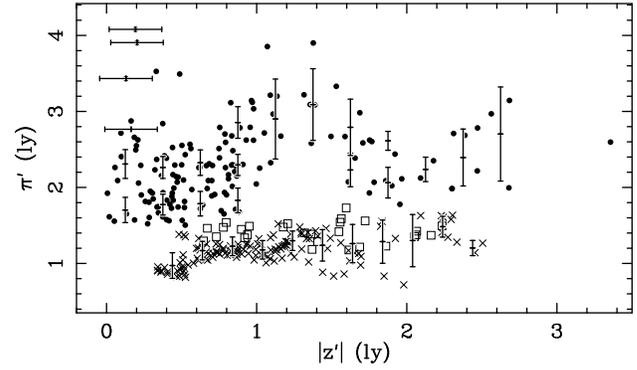}
\caption{As Fig.\ \ref{plCD4}, but showing the CS data.  Inner
  hourglass points (red in Fig.\ \ref{3dle}\pant{g--i}) are marked
  with ``$\times$'', extended-flux points (cyan and gold in Fig.\
  \ref{3dle}\pant{g--i}) with circles, and extended flux points
  belonging to the hourglass with squares.  The average radial
  profiles to each structure are overplotted.  Note that
  there are many binned averages in this plot, which are discussed in
  the text.
\label{plCS3}}
\end{figure}

As in the previous two subsections, we incline the data about this
axis, remove any ellipticity, and plot the cylindrical-radial profile
in Figure \ref{plCS3}.  Extended flux points that are now associated
with the hourglass are plotted as squares.  A large subset of the
extended-flux echoes (circles) are concentrated within 1 ly of the
equatorial plane $(z'=0)$, between 1.5--4 ly from the SN, and appear
to form a thick equatorial waist or belt circumscribing the hourglass.
As in the previous two sections, we measure the average radial profile
of these points in bins along $|z'|$.  There is much structure to
these data, which required the fitting of six profiles.  We again
visualize the nebula by reflecting each profile about the equator, and
revolving it around the cylindrical axis, as rendered in Figure
\ref{3dpl}\pant{i--p}, and sketched in cross section in Figure
\ref{toons}\pant{c}.  We identify five features in this data, discussed
below.

%\subsubsection{The CS Hourglass \label{sec-CS-shape-hourglass}}

The innermost radial structure is the hourglass, colored red in
Figure \ref{3dpl}\pant{i--p}.  This is fairly
cylindrical (opening angle $\sim$12\degr), with a semi-minor axis of 1.2
ly, tapering close to the equator and opening just slightly at large
$z'$.  A very small number of points suggest that it also tapers
around $z'=1.5-2$ ly and flares at $z'=0.9$ and 1.4 ly, but these
features are marginal.  Consistent with CKH95, we do not see evidence
for a ``capping surface'' to this structure, as would be expected from
a  peanut-like nebula.  To distinguish this hourglass from the
larger one containing the NH echoes, we refer to each by their
association, \ie the former is the ``CS hourglass.''

In Figure \ref{plCS3}, many of the extended-flux points (squares) lie
just outside the average position of the hourglass.  Since most of
these points are part of the southern lobe, this could suggest it is
wider than its northern counterpart.  A more likely interpretation is
that many of the southern-lobe points represent dust just outside the
densest part of the hourglass, which was itself illuminated prior to our
earliest epoch.

%\subsubsection{The Equatorial Plane and Belt 
%  \label{sec-CS-shape-belt}}

Immediately surrounding the CS hourglass is the thick waist of
extended high-surface brightness, which we call the ``belt'' (colored
green in Fig.\ \ref{3dpl}\pant{i--j}).  It extends $\pm1$ ly along the
$z'$ axis, and from 1.5 to 2.5 ly in radius.   Beyond the
outer radius of this belt, echoes lie fairly well constrained to the
equator, tracing a thinner (0.5 ly thickness in $z'$) but extended
equatorial plane (blue in Fig.\ \ref{3dpl}\pant{i--j}) to an outer
radius of $\pi'=4$ ly.  The belt does not appear to taper smoothly
into the thinner equatorial plane, but rather has a fairly sharp
transition at its outer boundary.

%\subsubsection{The Inner and Outer Walls  \label{sec-CS-shape-walls}}

The rest of the CS material lies along one of two ``walls.''  The
inner wall is colored cyan in Figure \ref{3dpl}\pant{i--j}, extending
1.5 to 2.6 ly along the axis, and may be considered to join with the
hourglass around $|z'|=1.5$ ly.  The outer wall, colored gold, bridges
the belt and the inner wall at radii between
2.6 and 3.4--4.0 ly.  We note the conspicuous lack of echo signal
between the hourglass and outer wall around $|z'|=1.4$ ly, and
question whether this is real, perhaps indicative of episodic
mass loss, or simply a gap in our data.

\subsection{The Geometry of the Outer Rings \label{sec-CS-rings}}

\begin{figure}\centering
\includegraphics[height=2.5in,angle=0]{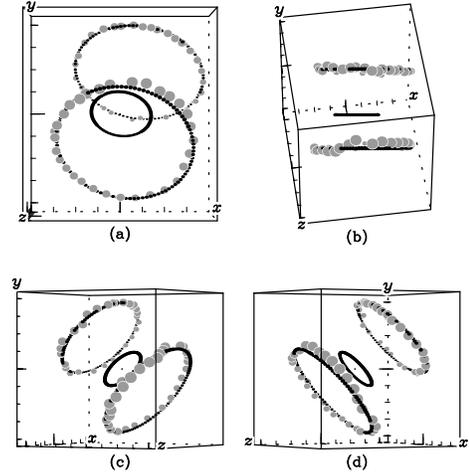}
\caption{Probable positions of the ORs (grey) and the planar,
  elliptical fits (black) to each ring, viewed \panp{a} face-on,
  \panp{b} 60\degr east, \panp{c} 60\degr west, and \panp{d}
  orthogonal to the inclination axis.
 \label{3de_ORs}}
\end{figure}

While great attention has been placed on the geometry of the ER
\citep{Jak91,Pla95,Bur95}, relatively little work exists on the ORs
\citep[see][]{Bur95,CH00}.  Now that a 3-D map of the
nebula exists, we can investigate how it can
constrains the geometry of the ORs (full details are in
in \sect{I-sec-CS-rings} of Paper \two).

The North OR (NOR) appears to lie along the outer edge of the north CS
hourglass, while the South OR (SOR) seems to lie at the
intersection between the southern hourglass and the belt, perhaps
indicating an interaction between wind that created the hourglass and
the pre-existing belt material.
 
The 3-D positions of the ORs are rendered in Figure \ref{3de_ORs}, and
appear reasonably planar.  We approximate both rings as planar
ellipses, fit to each distibution of points viewed along the CS
inclination axis.  The fitted parameters are listed in Table
\ref{tbl-CS-rings}, where $(x'_0,y'_0)$ are the centroid offsets from
the axis, $z'_0$ is the distance to the SN along the axis, $(a,b)$ are
the semi-major and minor axes, and $\phi$ is the \pa of the major axis
from north.  For completeness, the parameters for the ER
are also listed.  These fits are also indicated
in black in Figure \ref{3de_ORs}.  The approximation to the NOR
is quite good, but the SOR ellipse fails to intersect the points that
pass just north and west of the ER.  This is not suprising, since no
single ellipse can fully reproduce the observed shape of the SOR
unless it is non-planar. We have used these planar approximations
in the echo renderings throughout this work.

\begin{deluxetable}{l c c c c c c}
\tablewidth{0pt}
\tablecaption{Best-fit ellipses to the CS rings \label{tbl-CS-rings}}
\tablehead{
\colhead{Ring} & \colhead{$x'_0$} & \colhead{$y'_0$}  & 
  \colhead{$z'_0$}  & \colhead{$a$}  & \colhead{$b$} & \colhead{$\phi$} \\
\colhead{} & \colhead{(ly west)} & \colhead{(ly north)}  & 
  \colhead{(ly)}  & \colhead{(ly)}  & \colhead{(ly)} & \colhead{(\degr)}
}
\startdata
ER   & 0.015    & 0.0    &  0.0   & 0.647 & 0.98   & 81.1  \\
NOR  & 0.26     & 0.04   &  -1.36 & 1.42  & 0.94   & 70.5  \\ 
SOR  & 0.19     & 0.06   &  1.00  & 1.59  & 0.92   & -1.1 
\enddata
\end{deluxetable}

\begin{figure*}\centering
%\begin{figure}\centering
%\includegraphics[width=6.5in,angle=0]{toons}
\includegraphics[width=6.5in,angle=0]{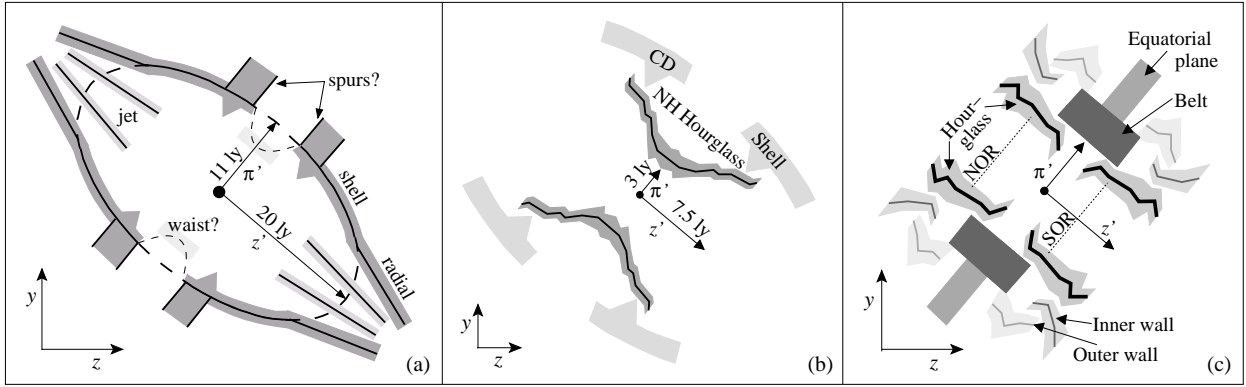}
\caption{Cartoon sketches of the salient structures traced out by
\panp{a} the CD model from Fig.\ \ref{plCD4}, \panp{b} the NH model
from Fig.\ \ref{plNH4}, and \panp{c} the CS model from Fig.\
\ref{plCS3}.  Figures are to
  scale, with the length of the two orientation arrows equal to
  \panp{a} 10 ly, \panp{b} 4 ly, and \panp{c} 2 ly.  Solid lines trace
  the radial profiles, and the widths of the greyscale regions trace
  the scatter of points about the averages.
\label{toons}}
%\end{figure}
\end{figure*}

\subsection{Summary}\label{sec-CSE-summary}

\begin{figure*}\centering
%\begin{figure}\centering
%\includegraphics[width=6in,angle=0]{3dpl}
\includegraphics[width=5.5in,angle=0]{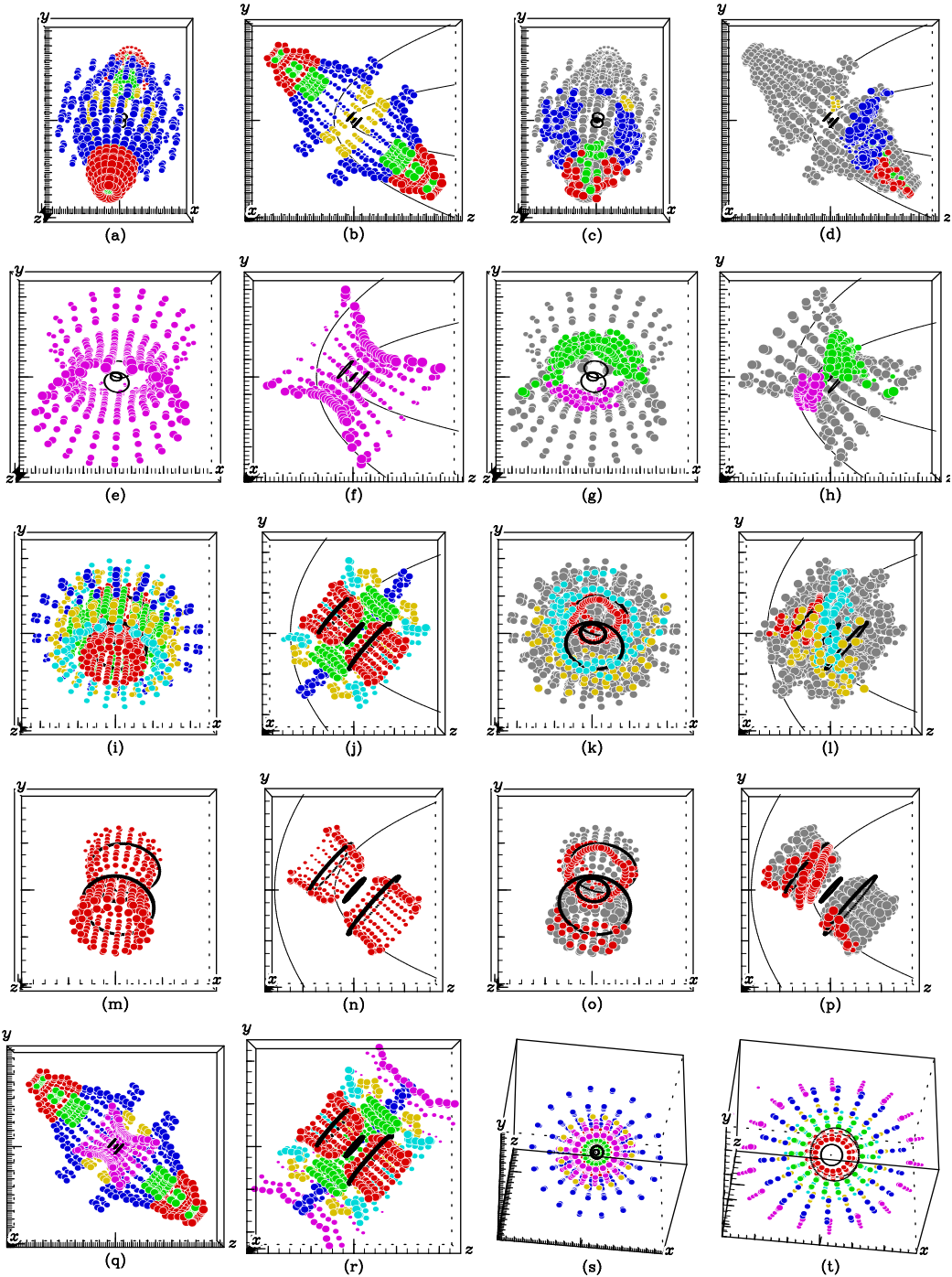}
\caption{Rendered views of the probable structures containing the
 observed echoes.  {\em Top Row}: The CD nebula.  {\em Second Row}:
 The NH nebula. {\em Third Row}: The CS nebula.  {\em Fourth Row}: The
 CS hourglass.  For these top four rows --- {\em Left column:} the
 face-on view of the complete structure; {\em Second column:} the
 western half, viewed from the east, showing a clear view of the
 interior; {\em Right columns:} face-on and side views of the complete
 structures in monotone grey, overlaid with actual echo points from
 Figure \ref{3dle}.  For the bottom row --- \panp{q} The western
 halves of the CD and NH, and \panp{r} NH and CS nebulae, viewed from
 the east.  \panp{s} The northern halves of the CD and NH, and
 \panp{t} the NH and CS nebulae, viewed along the inclination axes
 from the south.  See text for colors.
\label{3dpl}}
\end{figure*}
%\end{figure}

\begin{figure}\centering
\includegraphics[height=2in,angle=0]{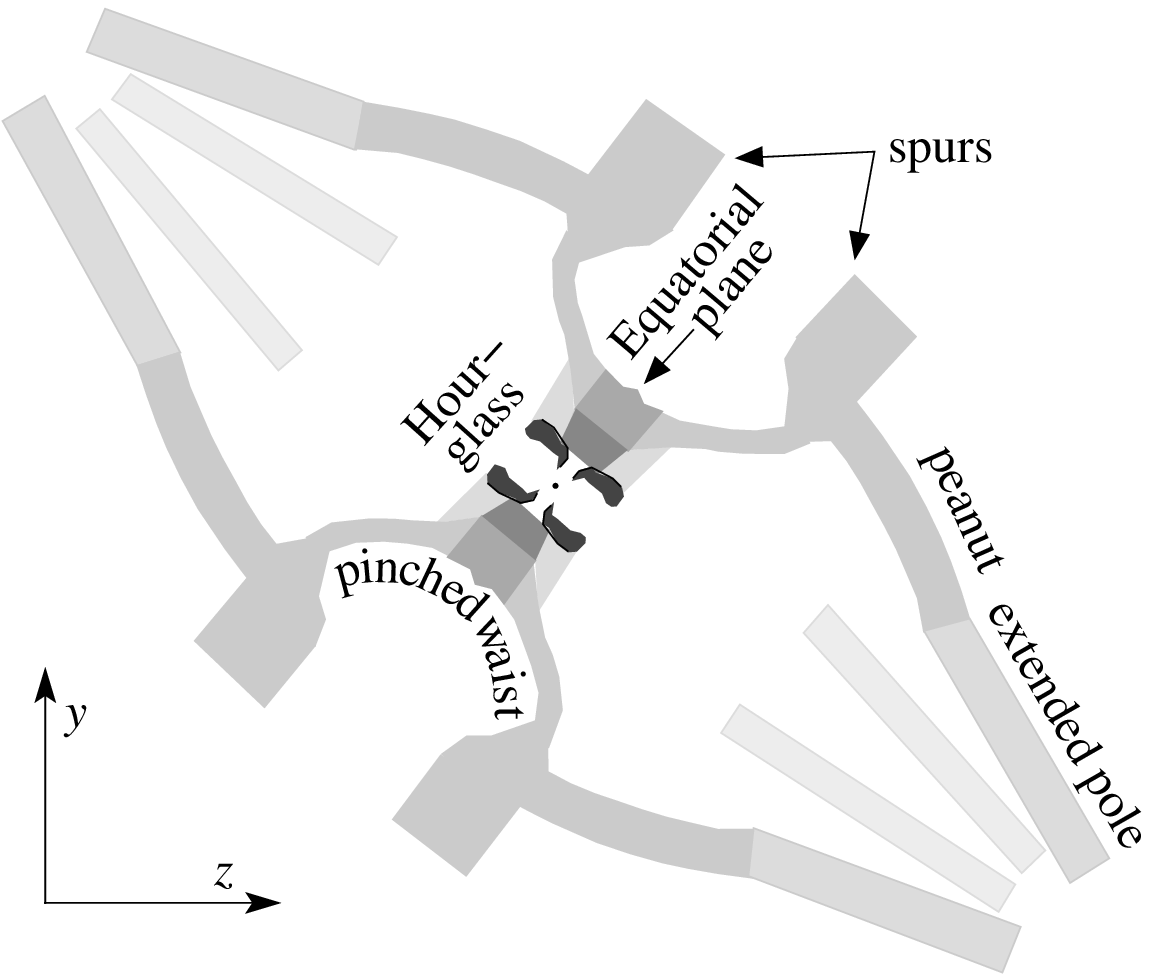}
\caption{Revised cartoon showing the simplified CSE suggested by
  combining the panels in Fig.\ \ref{toons}.  Structures are shaded to
  indicate density, which increases with greyscale.  Figure is to
  scale, and orientation arrows indicate 10 ly.
 \label{all_toon2}}
\end{figure}

The probable geometry and structure of the echoes within 25\arcsec of
SN~1987A are rendered in Figure \ref{3dpl}, and summarized
in Table \ref{tbl-density-geom}.  The final column gives the
approximate total volume of the structure containing each echo, found
by integrating the radial profiles in Figures \ref{plCD4},
\ref{plNH4}, and \ref{plCS3} along, and revolved $2\pi$ about, the
respective inclination axes.

%  The inner two structures
%(CS and NH) both have elliptical cross sections; although the
%ellipticities and orientations differ, it is unclear whether
%this difference is significant given .  

The east/west and north/south inclinations are quite consistent,
suggesting all circumstellar material shares a common inclination near
$i_x=40\degr$ south and $i_y=8\degr$ west.  Furthermore, the
structures appear to join to form a single nebula, as follows.  The CS
gas is nested neatly within the NH material, as shown in Figure
\ref{3dpl}\pant{r}, with the CS belt, equatorial plane, outer wall,
and the waist of NH consistent with a single, uniform thick waist,
extending inward to the CS hourglass (Fig.\ \ref{3dpl}\pant{o}).  The
outer edges of the NH hourglass almost reach the inner spur of the CD
shell (Fig.\ \ref{3dpl}\pant{q}), and indeed the echoes as viewed on
the sky do connect at early times (Fig.\ \ref{3dle}).  As shown in
Figure \ref{3dpl}\pant{m}, no material was illuminated along the
equatorial plane exterior to NH.\ \ We do not believe this region was
shadowed from the SN by material at smaller radii (\sect{sec-density}
of this paper, or \sect{I-sec-density-density-shadowing} of Paper
\two), thus it is unlikely there is any higher density gas outside the
waist of NH.  As such, we conclude the NH hourglass is the pinched
waist of the CD.

Figure \ref{all_toon2} shows all circumstellar material, simplified
according to the above arguments.  The outer structure is peanut-like,
extended along the poles to $r\lesssim28$ ly, and narrowly-pinched at
the waist at $r\sim5$ ly.  We will refer to this outer shell (traced
by the CD and outer NH echoes) as the ``Peanut.''  Roughly 35\degr
from the equatorial plane are the spurs, extending $3-4$ ly out from
each lobe, perpendicular to the Peanut's axis.  Unlike the
northwestern WFPC2 echoes, these spurs do appear at more than one
location, but they are limited enough in extent that it is unclear
whether they are uniform features that encircle the entire structure,
or isolated clumps.

Inside the equatorial region of the Peanut is a thick ($\sim 3.5$ ly
in radial thickness, 2 ly in axial length) annular ring (or belt),
which extends inward until it terminates at the inner CS hourglass.
The extent of this waist in $z'$ is unclear.  Echoes were detected
outside the CS hourglass at most equatorial ($\pi'$) radii between 1.6
and 4 ly, with a vertical ($z'$) distribution extending at least as
far as the CS hourglass itself.  Still, the bulk of material is within
1 ly of the equator.

\section{Density and Mass of the CSE }\label{sec-density}

%\begin{figure}\centering
\begin{figure*}\centering
\includegraphics[width=6.5in,angle=0]{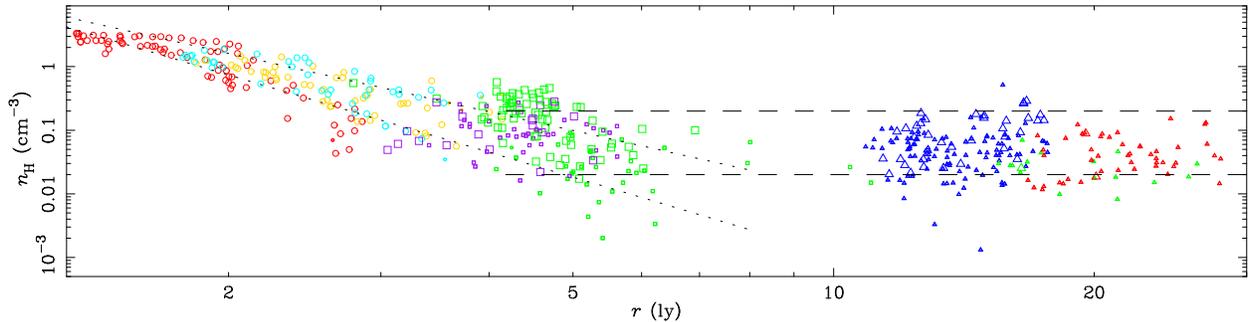}
\caption{ Average gas densities that match observed echo fluxes to the
 dust-scattering model.  Point colors correspond to those in the two
 right columns of Fig.\ \ref{3dpl}.  CS material is marked by circles,
 NH by squares, CD by triangles, and smaller symbols denote
 measurements with $S/N<2$.  Dotted lines are the best-fits through
 the upper and lower CS loci, corresponding to $n\propto r^{-3.1}$
 (upper) and $n\propto r^{-4.5}$ (lower).  Horizontal dashed lines
 delineate the rough density boundaries of the CD gas.
\label{enh}}
\end{figure*}
%\end{figure}

To constrain the gas density and dust composition of the CSE, we apply
a dust-scattering model (\sect{I-sec-le} of Paper \one) using the
surface brightnesses of all echoes discussed in the previous section.
The reader is referred to \sect{I-sec-density} of Paper \two for the
complete analysis, as only a brief summary is presented here.

The CD echoes are best fit with large ($a=3.5$\AA$ -
2.0$\micron) grains consistent with an LMC-abundance of silicate and
carbonaceous dust.  Outer NH data ($r>5$ ly) are more consistent with
Si-dominated large grains, but perhaps with a slightly-smaller maximum
grain size than for the CD.  The inner NH points, along with the CS
echoes, are well fit by Si-only dust with $a_{max}=0.2$\micron.  The
general trend is that inner echoes are better reproduced by smaller,
Si-dominated dust, and with increasing distance, the grain sizes and
C-content increases.

\citet{Fis02} have modeled mid-IR emission from collisionally-excited
dust grains heated in the shocks between the CS gas and the forward
blast of the SN.  They find the emission is best explained by small
($a\lesssim0.25\micron$) grains with a Si-Fe or Si-C composition.  The
dust abundance is quite low, which they attribute to evaporation from
the UV flash and sputtering in the shocked gas.  This further
constrains grain sizes to $a\lesssim0.25\micron$ and excludes a
pure-carbon composition.  The dust in both this shocked region and the
CS hourglass was formed from material expelled very late in the
progenitor's life, thus we expect the dust properties to be similar.
Indeed, our pre-SN CS dust model also favors silicate-dominated dust
with grain sizes $a<0.2\micron$.

That dust which formed at later times has a smaller carbonaceous
content can result from a change in surface CNO abundances over the
star's late-stages of evolution, since carbon-rich envelopes create
carbon-rich dust, while oxygen-rich envelopes create silicate-rich
dust.  Such CNO processing is also inferred from early IUE spectra
\citep{Fra89} of the ER, which show nitrogen and oxygen to be
overabundant with respect to carbon.

The average dust density is plotted in Figure \ref{enh}.  Figure
\ref{all_toon2} has been greyscale shaded to reflect the differences
in density among the many echoing structures, with darker grey
indicating higher-density material.

The innermost CS hourglass material has a relatively-constant gas
density of $n_{\rm H}=2-3$ cm$^{-3}$ up to $r\sim1.6$ ly.  Beyond this
position, the material splits into two distributions, where the
steeper profile ($r^{-4.5}$) traces the denser waist of the hourglass.
Beyond $r=2$ ly, the shallower ($r^{-3}$) profile traces the belt and
equatorial plane material, making a smooth transition to the inner NH
points around $r\gtrsim 3$ ly.   That the CS density varies more steeply
than the inverse-square expected for a freely-expanding wind suggests
the mass-loss mechanical luminosity ($\dot{M} v_{exp}^2$) increased
with time toward the end of the RSG.

Excluding a density enhancement between $r=4-5$ ly, there is no
evidence of a structural distinction between the CS and NH gas, which
justifies the simplified model in Figure \ref{all_toon2}.  Returning
to the aforementioned enhancement, this higher-density material is
located along the waist of the NH hourglass, marking an outer edge to
the equatorial overdensity.  The constituents of the Peanut, the CD
and outer NH, have constant density with radius, bounded by $0.02\le
n_{\rm H} \le 0.2$ cm$^{-3}$, suggesting this structure marks a
constant-density boundary.

\citet{Bur95} argued that {\em if} a CS hourglass exists, the fact
that it is not observed in recombination (like the three rings)
implies its density must be $\lesssim5$ cm$^{-3}$.  This density is
consistent with our findings, and explains why the hourglass exists
but is optically invisible.  

In \sect{I-sec-density-density-unbiased} of Paper \two, we present
an unbiased measurement of density by searching for echoes at all
positions within each difference image.  These reveal the region
between the CS and CD structures to be filled with diffuse material
($n_{\rm H}\lesssim0.03$), while little structure is seen outside the
CD.  According to Figure \ref{all_toon2}, the CD and NH form a peanut
with a narrowly-pinched waist.  If the equatorial dust density is high
enough to cause significant extinction, material beyond NH will be
effectively shadowed from the SN light pulse, in which case no echoes
would be seen.  In \sect{I-sec-density-density-shadowing} of Paper \two,
we show that only a fraction of a magnitude of extinction can be
expected from known equatorial material.  As this is insufficient to
cause pronounced shadowing, the narrow waist appears to be a genuine
boundary.

\begin{deluxetable*}{l c c c  c c c  c c c c c c c }
\tablecaption{Summary of Echo Geometries \label{tbl-density-geom}}
\tablewidth{0pt}
\tablehead{
\colhead{Struc-} &
 \multicolumn{3}{c}{Inclination\tablenotemark{a}} &
 \multicolumn{3}{c}{Cross Section\tablenotemark{b}} &
 \multicolumn{1}{c}{} &
 \multicolumn{5}{c}{Geometry\tablenotemark{c}}
\\
\colhead{ture} &
 \colhead{South} & \colhead{East} & \colhead{Roll} &
 \colhead{$b/a$} & \colhead{$\Delta x_0$} & \colhead{$\Delta y_0$} &
 \colhead{Shape} & \colhead{$r$} & \colhead{$z'$} & \colhead{$\pi'$} &
    \colhead{$\Delta\pi'$} & \colhead{$V_{tot}$}
\\
\colhead{} &
 \colhead{(\degr)} & \colhead{(\degr)} & \colhead{(\degr)} &
 \colhead{} & \colhead{(ly)} & \colhead{(ly)} &
 \colhead{} & \colhead{(ly)} & \colhead{(ly)} & \colhead{(ly)} &
    \colhead{(ly)} & \colhead{(ly$^3$)}
}
\startdata
%  % i_x     & i_y     & i_z    & b/a    & Delta_x  & Delta_y   &
%  shape     &  r     &  z      &  pi     &  dpi     &  Vobs    & Vtot
CS & 41\degr & -8\degr & -9\degr & 0.94 & $\lesssim 0.1$ & 0. &
   Hourglass & $1-2.9$ & $0.3-2.5$ & $0.8-1.6$ & $0.5-1$ & 20  \\
   &         &         &        &      &                &    &
   Belt      & $1.5-2.8$ & $0.-1.0 $ & $1.5-2.6$ & $1.5-2.0$ & 40  \\
   &         &         &        &      &                &    &
   Walls     & $1.8-4.2$ & $0.8-2.7$ & $1.8-3.6$ & $2.0-3.0$ & 120  \\
NH & 40\degr & -7\degr & 12\degr & 0.82 &  0. & 0. &
   Hourglass & $3-11$  & $0.7-7.5$ & $2.8-7.5$ & $2-4$ & $1300$ & \\
CD & 40\degr & -8\degr & \nodata & $\gtrsim 0.95$ & $<-1$ &
   $\lesssim -0.1$ &
   Prolate   & $10-28$ & $4-28$ & $0-16$ & $4-5$ & $1.2\times10^4$
\enddata
\tablenotetext{a}{``Roll'' is the counter-clockwise rotation about the
   $z$ axis, also the P.A.\ of the minor axis after removing the
   southern inclination and eastern rotation. }
\tablenotetext{b}{$\Delta x_0$ and $\Delta y_0$ are the western and
   northern offsets of the symmetry axis from the SN, which is at
   $x=0$, $y=0$.}
\tablenotetext{c}{$r$ is the spherical-polar distance from the
   SN. $\pi'$ is the cylindrical-polar radius measured from the
   inclined axis $z'$.  
 $\Delta\pi'$ is the approximate width along $\pi'$.
 $V_{tot}$ is the inferred total volume of the structure.}
\end{deluxetable*}

The total mass of gas and dust within the CSE is estimated from the
average density and volume of each structure, as given in Figure
\ref{enh} and Table \ref{tbl-density-geom}.  The volume interior to
the CD is estimated a few ways: as the volume of the ellipsoid
enclosed by the CD, as the volume of two cones enclosed by the
quasi-linear profile, and as the integrated volume under the same
profile (Fig.\ \ref{plCD4}).  All three methods give a consistent
result of about 6500 ly$^3$. The calculated masses are listed in Table
\ref{tbl-density-mass}, yielding a total mass in all structures of
$\sim 1.7$\msun (assuming an LMC gas-to-dust mass ratio of
400--600). It is very difficult to quantify all the sources of error,
but we believe the tendency will be to underpredict the masses of the
structures, given conservative measurements of their volume and the
unknown extent to which dust was not observable, due to unfavorable
geometry, poor data quality, confusion, or possible shadowing.

\section{Previous Formation Models}\label{sec-echointerp}

As discussed briefly in \sect{sec-intro}, the progenitor's
pre-explosion spectral type, as well as the early SN evolution, argue
Sk $-69$\degr~202 exploded as a BSG.  Evidence of pre-existing CS
material, and nitrogen enrichment within that material, imply the star
first passed through a RSG phase, executing a ``blue loop'' to return
to a BSG prior to core collapse.  

Mass loss can explain this blue loop, however only if the star
loses either a small fraction or nearly all of its hydrogen envelope
\citep{WPE88,WPW88,Nom88}.  While the latter option is ruled out by
observations \citep{Dop88,Wal89}, even low mass loss can 
qualitatively account for the observed CSE via interacting stellar
winds \citep[or ISW,][]{Kwo82,Bal87}.  Originally proposed to explain
planetary nebulae, a slow, dense wind from the evolved star (or RSG)
is overtaken by a fast, tenuous one, accelerated from the hot, blue
core (or BSG).  In particular, inhomogeneities in the winds, such as
equatorial overdensities, shock focus the winds into characteristic
bipolar shapes.

Unfortunately, the formation mechanisms of such overdensities are
poorly understood.  \citet{Mor81,Mor87} proposed that a binary
companion can create an overdensity in the primary wind within the
pair's orbital plane, while \citet{SL94} argue that this configuration
can drive high-velocity jets out of that plane.  Two models exist for
the creation of a disk by single stars.  A rotation-induced
bi-stability \citep{LP91} produces a larger mass-loss or lower
wind-velocity at the equator than the poles, resulting from the high
sensitivity of line-driven winds to optical depth.  \citet{BC93} have
introduced the model of a wind-compressed disk (WCD), in which the
orbital streamlines of gas launched from the northern and southern
hemispheres of a star collide, producing an equatorial overdensity.

Many of these mechanisms have been invoked to explain the formation of
the three-ring nebula, which has been the most prominent feature of
the CSE.  In the following subsections, we review these models, and
compare their predictions to the complete nebula revealed by light
echoes.  In short, we find that no extant model can adequately
reproduce this system, however a number of features among them may
offer some insight into the progentor's probable mass-loss history.

\subsection{Interacting Stellar Winds Models}\label{sec-echointerp-ISW-ISW}

\citet{Luo91} first invoked the ISW model to explain the emerging
picture of an hourglass-shaped nebula containing the three rings.
Using the thin-shell approximation \citep{MM88} in which all swept-up
material lies in a thin shell with an isobaric interior (effectively a
momentum-conserving snowplow), they simulated the collision of a fast,
tenuous BSG wind with an older, slow and dense RSG outflow.  Since,
this collision has been studied with 2-D hydrodynamic simulations by
\citet{BL93} and \citet{MA95}.

A number of results are similar between all of these models. (1) A
bipolar or hourglass nebula can be formed via this ISW framework that
roughly matches the size and expansion velocities of the rings.  (2)
To match these constraints, an extraordinary asymmetry is required in
the RSG mass loss, producing a significant equatorial density
enhancement over that at the poles.  We note that this conclusion is a
result of the simulations, not an input condition.  (3) The wind
parameters necessary to produce this nebula are consistent with
average values of other stars, with the RSG characterized by a mass
loss $\dot{M_R}=6\times10^{-6}-10^{-5}$ \msunyr, wind velocity
$v_R=8-10$ \kms, and lifetime $\tau_R<10^6$ yr, and the BSG by
$\dot{M_B}=1.5\times10^{-7}-3\times10^{-6}$ \msunyr, $v_B=300-550$
\kms, and $\tau_B=10^4$ yr.

As predicted above, we do find an extended equatorial overdensity in
Figure \ref{all_toon2}.  However, compared to this newly-revealed
CSE, the models all have significant shortcomings.  Foremost, nearly
all the predicted CS nebulae 
are quite flattened and oblate (mushroom shaped), inconsistent
with the prolate, hourglass shape we have mapped out.  To generate
roughly the right shape, the
equatorial expansion velocities become far too large, while matching
these velocities can lead to RSG mass-loss rates that are too high.
Although these models can explain the formation of the ER, they fail
to explain the existence of the ORs.  This may be linked to the fact
that almost all model hourglass nebulae have solid capping surfaces,
which we do not observe.  Finally, the adoption of a pre-existing RSG
asymmetry is entirely {\em ad hoc}, and these models make no attempt to
explain {\em how} such an asymmetry formed, nor to model what
geometry its interface with the MS bubble should be.

%It is unclear from their description whether the MA95
%models followed the stellar evolutionary sequence presented in the
%same paper (see \sect{sec-echointerp-evol}).  In agreement
%with \citet{LEB89}, their evolutionary calculations suggest the star
%underwent two blue loops, with roughly equal time spent in the
%red and blue stages, but with the first loop roughly 10 times longer
%than the second.  MA95 may have presented results from
%early in the {\em first} BSG loop, in which case it is unknown from
%their work what effect a prolonged BSG wind, followed by a
%short-duration RSG-to-BSG loop, would have on the CSM and the
%resulting structure.

\subsection{A Wind-Compressed Disk Model}\label{sec-echointerp-ISW-WCD}

\citet{Col99} have attempted to address one shortcoming in previous
work by assuming that the RSG wind forms a WCD from which the density
asymmetry is derived.  In contrast to the above work, they also allow
for a velocity asymmetry in the RSG wind, as defined by the disk.  A
WCD is characterized by the rotation parameter
$\Omega=V_{rot}/V_{crit}$, the ratio of the stellar rotation rate to
the break-up speed.  For rotations less than a threshold value
$\Omega<\Omega_{th}$, the star forms a wind-compressed {\em zone} with
an equator-to-pole density contrast of 3--10, while for
$\Omega>\Omega_{th}$ a WCD forms with density contrasts $\ssim100$.
By varying the WCD parameters, \citet{Col99} varied the RSG-wind
environment surrounding the star, establishing the initial conditions
for a BSG wind.  This they addressed using self-similar solutions of
wind-blown bubbles from \citet{DCB96}.

For their fiducial model (a RSG with 20\msun, which is too high from
stellar-evolution arguments), they find $\Omega_{th}\sim0.3$, which
produces a density contrast of 25.  A variation of the rotation
parameter between $\Omega=0.3-0.36$ produces peanut-shaped nebulae
with equatorial expansion velocities of 8--10 \kms and a maximum
(cylindrical) lobe radius roughly twice that in the equator
(consistent with the diameters of the ORs and ER), for RSG wind
parameters $\dot{M}_R\sim10^{-5}$ \msunyr and $v_R=20$ \kms, and BSG
parameters $\dot{M}_B\sim10^{-7}$ \msunyr and $v_B=400$ \kms.  A WCD
is not required to produce a peanut, as a dense equatorial zone is
sufficient.  Again, the peanut shape does not fit the observed CS
hourglass, however they address this by noting that in simulations
\citep{FM94}, the BSG wind can be shock-focused into the polar
direction, resulting in a substantially-more prolate structure than
that predicted from a self-similar solution.

A few results stand out from this work.  (1) The WCD is a
straightforward explanation for the observed equatorial density
enhancement.  (2) The authors do not find the need to invoke unusually
low BSG wind momentum or RSG velocity, although the RSG mass-loss rate
is much higher than we deduce in \sect{sec-evol-RSG}.  (3) A single
RSG star cannot rotate at $\Omega\gtrsim0.25$ without the influence of
a binary companion.  \citet{Col99} estimate that a lower limit of
0.6\msun for the companion is sufficient to produce the necessary
rotation speeds for this model.  If rotation is to be invoked as a
formation mechanism, then the progenitor had to have been a binary
system.

\subsection{Binary Models}\label{sec-echointerp-ISW-binary}

Binarity has been invoked by \citet{Pod89} and \citet{Pod90} among
others to explain the observed characteristics of the SN.  A companion
could have been engulfed by Sk $-69$\degr 202 during a common-envelope
phase; however, for lower mass primaries, the secondary can cause
dynamical instabilities that eject the primary envelope
\citep{RL96}, a process that is not suggested from the SN evolution.

\citet{Sok99} invokes the influence of a binary companion to produce
the observed three-ring nebula, but his results are inconsistent with
the ensemble of data presented here.  \citet{Pod91} suggest that the
collision of winds between the primary and secondary spread out the
wind into a truncated double cone.  This is a novel formation
mechanism for the hourglass feature we observe; however, their
geometric model further predicts that Napoleon's Hat lies on the
northern-half of the cone, which we now know to be false.

\citet{LOK95} suggest a variant on the double-cone scenario, in which
the BSG wind sweeps up the base of the cones, leaving the end surfaces
as the ORs.  Furthermore, dynamical friction causes the secondary to
coalesce into the primary, ejecting the envelope, thereby forming the
ER.  Again, this model is attractive as it suggests that a single
mechanism formed all three rings.  A testable prediction is that
extensive RSG double-cone material should exist at and beyond the ring
positions, however we see no evidence for such material in light
echoes.  If binarity is to explain the formation of the three rings,
these models must be revisited to be consistent with the new picture of
the complete CSE.

\subsection{Ionization}\label{sec-echointerp-ISW-ionization}

All the above
models have neglected the effects of ionization on the evolution of
ISWs.  The progenitor MS star is expected to emit an ionizing flux of
$S=2\times10^{48}\; \gamma$ s$^{-1}$, while the B3 I BSG star should
emit $S=4\times10^{45}\; \gamma$ s$^{-1}$ \citep{Pan73}, both of which
are sufficient to drive an ionization front ahead of the outer shock.

\citet{CD95} have addressed the role of ionization in the to explain
the reappearance of radio emission three years after outburst
\citep{Man01}.  Their model of a pre-SN \ion{H}{2} region suggests
that the BSG had a mass loss of only $\dot{M}_B=8\times10^{-8}$
\msunyr with $v_B=450$ \kms.  Furthermore, the ionization front that
formed this \ion{H}{2} region would move rapidly through the
lower-density polar regions of the hourglass, and could have broken
out of the lobes, thereby eliminating any capping-surface to the CS
hourglass.  This is a particularly interesting scenario for explaining
the observed geometry of the CS hourglass, as well as the evolution of
the SN remnant across the spectrum.

\citet{Mey97} has proposed that an ionization front in the BSG phase
induces hydrodynamic motions in the pre-existing asymmetric RSG wind.
These motions create a latitudinally-dependent density profile that
peaks about 50\degr above the equator, which he identifies as the
outer rings, however \citet{Sok99} and \citet{CH00} have offered a few
objections to this scenario.  Fully-radiative hydrodynamic simulations
including ionization are necessary to test these ideas.

\section{Constraints on the Evolution of Sk $-69$\degr 202}\label{sec-evol}  

We now investigate what constraints the geometry and density of the
CSE can place on the progenitor's evolution.

\subsection{The Main Sequence Bubble \label{sec-evol-MSbubble}}

We begin by discussing the possible mass-loss scenarios for the MS
progenitor, since the impact of these winds on the surrounding ISM
forms the initial conditions for subsequent mass-loss evolution.
\citet{CE89} note that the wind should have blown a bubble $>10$ pc in
radius, whose interior homogenized as the star evolved onto the RSG
branch.

Let us assume that Sk $-69$\degr 202 was a zero-age MS type O9 V with
18--20\msun \citep{CE89}, luminosity $10^{4.9}L_\sun$ and radius
$8R_\sun$ \citep{Pan73}, yielding a MS mass-loss rate of
$\dot{M}_{MS}\sim10^{-7}$ \msunyr \citep{NdJ90}.  The interaction of a
long-lived stellar wind with the ISM was first investigated by
\citet{CMW75} and \citet{Wea77}, with refinements proposed by, among
others, \citet{McK84}, Koo \& McKee (1992a,b), and \citet{DEr92}.  The
wind is characterized by its mechanical luminosity
$L_w=\frac{1}{2}\dot{M}v_{\infty}^2$, where the terminal velocity has
an upper limit around $v_{\infty}\lesssim1500$ \kms for the assumed
progenitor properties \citep{Kud89,LC99}.  These yield $L_{36}=0.07$,
where $L_{36}=L_w/10^{36}$ ergs~s$^{-1}$.  \citet{WPW88} suggest the
mass loss rate may be higher, or $L_{36}\lesssim0.2$.

For the wind duration, we use a MS lifetime of $t_{MS}=8$ Myr \citep{MA95}.
\citet{Scu96} find $N($\ion{H}{2}$)= 4\times10^{21}$ cm$^2$ along the
line of
sight to Star 2, with roughly 30\% of the extinction occuring within
the LMC.  \citet{Xu95} find that SN~1987A is positioned roughly 1 kpc
behind the edge of the LMC, and if one assumes Star 2 is near the SN,
these yield an average ISM density of $n_0$=0.4 cm$^{-3}$.

For $L_{36}=0.07-0.2$, the progenitor wind will
maintain an adiabatic shock for 0.3--1.3 Myr.  As this is
much shorter than $t_{MS}$, the bubble will become radiative at a
radius of 27--81 ly.  Assuming the sound speed of the undisturbed ISM
is 10 \kms, the radiative shock front will reach this speed and stall
in another $\sim$0.8 Myr.  Following the arguments in \citet{DEr92},
adiabatic shocks become radiative at radii larger than those at which
radiative shocks would have stalled.  As such, the shock must recede,
stalling at a new radius of 26-42 ly.

There are subtle indications in our density measurements of an
enhancement 80--100 ly from the SN, which may be the remnant of the
MS-ISM snowplow.  However, given the small expected mechanical
luminosity of the progenitor, we question whether the Peanut is
actually the remnant of that interaction.  Alternatively, the MS
bubble may be traced by echoes seen roughly 100 pc from the SN by
\citet{Xu95}.

\subsection{Red Supergiant Mass Loss \label{sec-evol-RSG}}

\begin{deluxetable}{c c}
\tablewidth{0 pt}
\tablecaption{Mass of the CSM \label{tbl-density-mass}}
\tablehead{
\colhead{Structure} & \colhead{$M/M_\sun$}
}
\startdata
CS Hourglass & 0.04 \\
CS Belt  & 0.04 \\
CS Walls & 0.06 \\
NH Waist & 0.06 \\ 
NH Walls & 0.07 \\
Intra-CD & 0.2 \\
CD       & 1.2  \\ \hline
Total    & 1.7 
\enddata
\end{deluxetable}

Following the end of the MS, the gas density within its equilibrated
bubble is expected to be quite low, and for a large RSG mass-loss
rate, the swept-up mass within the driven shock will be much lower
than the mass of shocked wind.  The RSG mass-loss rate can be
estimated from the mass contained within this structure
($\sim$1.6 \msun, Table \ref{tbl-density-mass}) and an adopted RSG
lifetime of $3\times10^5$ yr \citep{MA95}, yielding
$\dot{M}_R\sim5\times10^{-6}$ \msunyr, consistent with ISW models.  A
wind velocity of 10 \kms corresponds to a mechanical luminosity of
$L_{36}=1.6\times10^{-4}$ which, expanding into a tenuous medium, is
classified by \citet{Koo92a} as a ``slow wind'' that becomes radiative
almost immediately.  Note that this rate is an upper limit if the MS
bubble is small, as suggested above.

\subsection{Inhomogeneities}

If the Peanut is the RSG-MS contact discontinuity, then its shape is a
fossil record of the hydrodynamic interaction between these two media.
Here, we investigate whether the observed Peanut can be explained by a
simple asymmetry in the RSG wind, or whether external factors need to
be invoked.  Consider the CD shell (Fig.\ \ref{toons}\pant{a}, which
has an elongated, prolate shape, closest to the SN just above the
equatorial plane ($r\sim11$ ly), and farthest toward the poles
($r\lesssim28$ ly).  A slow, radiative shock propagates as
$R(t)\propto L^{1/4}t^{1/2}$, thus for the RSG wind to have formed
this pinched structure in the absence of any external inhomogeneities,
the wind luminosity above the equator must have been roughly 3\% that
at the pole.

\citet{DEr92} find that radiative shocks stall at $R_m=22.2
L_{36}^{1/2}n_0^{-1/2}$.  For the RSG wind from Sk $-69$\degr 202 to
have stalled at the poles with $L_{36}=1.6\times10^{-4}$, the ambient
MS-bubble density must be $n_0=10^{-4}$ cm$^{-3}$.  Similar average
values can be determined by considering the MS mass-loss rate and
bubble size.  With 3\% of the polar wind luminosity, the
equatorial wind would have stalled at 5 ly, which is less than half
the observed position of the CD shell.  It appears that the
observed geometry of the Peanut cannot be explained by only an
asymmetry in the RSG wind.

One solution is that the CSM in the MS-bubble is not homogenous, but
that the MS winds had an asymmetry similar to that of the RSG.
Typically, one argues that the average temperature inside the MS
bubble should have been sufficiently high for any inhomogeneities to
smooth out, making this scenario unlikely.  An asymmetric
freely-expanding flow into a homogenous post-shock medium will create
an asymmtric inner termination shock.  The simulations of
\citet{GLM96b} show that this shock collapses into RSG wind before the
system settles anew into a steady flow.  Thus it may be possible for a
MS-wind asymmetry to be imprinted on subsequent outflows.  An
alternative is that the CD had a smaller size and asymmetry at the end
of the RSG, and a second wind blew it into the shape currently
observed, which we consider in greater detail in the next subsection.

\subsection{A Two-Loop Scenario}

\begin{figure}\centering
\includegraphics[width=3.in,angle=0]{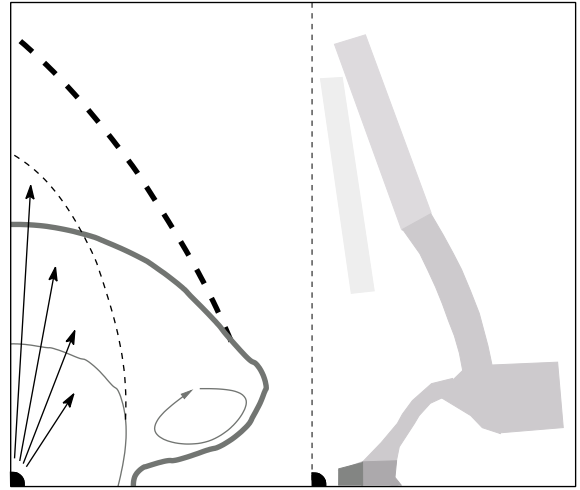}
\caption{Schematic showing a qualitative formation scenario for the
Peanut.  At left, the grey curves trace the favored 
model of BSG/RSG interaction from \citet{MA95}.  Pressure or momentum
directed toward the poles (arrows) blow this mushroom into a peanut
shape (black dotted lines).  At right, one quadrant of the simplified
Peanut from Fig.\ \ref{all_toon2}.  
\label{newISW}}
\end{figure}

\citet{MA95} have modelled the
 evolution of metal poor stars between 19--21 \msun using
the stellar evolution code from \citet{Arn91}.  All of these 
make one full blue loop and then return to the red, with only the
models under 20.5 \msun finishing a {\em second} loop to end as a BSG.
Such stars spend roughly $10^7$ yr in the MS, $2.3\times10^5$ and
$3.2\times10^5$ yr in the first RSG and BSG, and $4.7\times10^4$ and
$2.4\times10^4$ yr in the second RSG and BSG, respectively.  
We propose that the structures revealed by light echoes may be
explained by such an evolutionary scenario.  In particular, 
a long-duration BSG wind expanding into an asymmetric RSG
environment could be the second wind hypothesized above.

Qualitatively, the hourglass-shaped bubbles resulting from ISW
simulations are poor matches to the CS hourglass, yet they have many
similarities to the Peanut (Fig.\ \ref{newISW}).  The flattened,
oblate (or mushroom-shaped) lobes with narrowly-pinched waists from
simulations are very similar to the inner walls and conical-shell
contributions to the Peanut.  This is shown schematically in Figure
\ref{newISW}, in which the inward-facing shock (thin grey curve),
dense contact discontinuity (thick grey curve) and vortex (grey
arrowed curve) from Figure 2 of \citet{MA95} are shown at left, and
compared on the right to one quadrant of the CD structure from Figure
\ref{all_toon2}. Of note in \citet{MA95}, a dense vortex appears at
the outer edge of the prolate lobe, which is tempting to identify with
the ``spurs'' located just outside the CD.

We envision the following evolutionary sequence.  A BSG accelerates
wind into previously-expelled RSG material, creating a mushroom-shaped
discontinuity.  This BSG wind is steady and long-lived, and the
interface expands self-similarly \citep{DCB96} to the size of the
CD/NH Peanut.  Many mechanisms exist to direct momentum and pressure
toward the polar region (indicated by black arrows in Fig.\
\ref{newISW}), elongating the mushroom into the prolate peanut shape
we observe.  These include: (1) shock focusing of the BSG wind by
equatorial material \citep[see][]{Col99}, and preferential expansion
into the lower density polar regions of (2) an adiabatic shock or (3)
an ionization front \citep{CD95}. The star then loops back to the red,
and for a few$\times10^4$ years fills in some fraction of the cavity
blown out by the first BSG wind.  These three mass-loss phases produce
an extensive equatorial overdensity, which directs a
substantial amount of the final BSG wind in the polar direction,
carving out an hourglass-shaped nebula with no end caps.

One major discrepancy between the ISW models and the CSE as revealed
through echoes is that the models assume there is a very extensive
equatorial disk of RSG material.  Whether the equatorial material we
do see is sufficient to form the rest of the bipolar structures must
be tested in hydrodynamic simulations. 

In addition to providing a natural mechanism for formation of the
equatorial overdensity via a WCD, stellar rotation may also explain
the progenitor's evolutionary blue loop(s) \citep{Mey04}, and should
be revisited with hydrodynamic models.  From Figure \ref{enh}, we find
a maximum equator-to-pole density contrast of 10 between the NH
equator and the furthest CD material.  This is small compared to
values found by \citet{Col99}, suggesting the evolved star may only
have produced a wind-compressed zone and therefore did not have an
extremely high rotation rate.  However, since a rotating star spins
down as it loses mass on the MS and as it swells into a RSG, 
any significant RSG rotation requires some mechanism for spinning up
the star, such as a close binary companion.

\section{Conclusion \label{sec-conclusion}}

Once it was realized that Sk -69\degr 202 was surrounded by an
extensive CS nebula, it became clear that the closest observed SN in
400 years would also provide the first opportinity to observe the
destruction of that nebula \cite[e.g.][]{LM91a,BBM97b}, and the birth
of a supernova remnant (SNR), both spatially resolved and in
real-time.  This event thus serves as a vital test to models of SN
evolution, radiative shock processes, and SNR formation.

As noted in the introduction, the ejecta-CSE interaction will directly
illuminate the structures it impacts, and the high-energy emission
from the resulting hot gas will illuminate many other regions of the
CSE via photoionization.  The results from this work will be tested by
this interaction, and be critical in interpreting the panchromatic
spectacle that will ensue.

Already, the ``hot spots'' detailing the impact of the ejecta with the
inner edge of the ER \citep{LSB00} confirm the observed spatial offset
between the SN and inner nebulae.  \citet{Sug02} show that roughly
$\frac{3}{4}$ of the hot spots are located along the eastern half of
the ER, while spots toward that half of the ring are also closer to
the SN than those to the west.  Both of these are suggestive of an
ER that is offset to the west of the SN, consistent with our findings
in \sect{sec-CS}.  

An alternative interpretation to the hot spot positions and order of
appearance is that the circumstellar medium into which the ejecta
propagate is asymmetrically denser to the east \citep{Mic01}.  Is such
an imprint visible in light-echo data?  

Stellar winds propagating into uniform media travel faster and farther
with increasing mechanical luminosity $\rho v^2$ \citep{Wea77}.  A
higher wind density to the east would imply that the CD boundary
separating the RSG wind from the MS bubble should be more distant from
the SN in that direction.  In \sect{sec-CD}, we find that the CD does
appear marginally offset from the SN toward the east.  Although by no
means conclusive, this is suggestive that an eastern asymmetry was
present in the stellar winds as early as the RSG.

If the RSG outflows had been denser toward the east, then the inner
cavity which the BSG wind carved out would be closer to the central
star in that direction as well.  This would have resulted in an
observed ER and CS hourglass that are offset west of the SN, and an
overall density enhancement within the equatorial plane to the east.
This qualitative argument is again consistent with the observed
geometry of the inner CS structures \secp{sec-CS}.  Unfortunately, the
inferred densities from \sect{sec-density} are too uncertain to permit
a meaningful study of azimuthal asymmetries.  

In keeping with the binary-star hypotheses to explain the observed
CSE, one could also invoke binarity as the cause for the east-west
wind asymmetry.  For example, a secondary in a resonant, eccentric
orbit could tidally distort the primary when near pericenter, and from
the von Zeipel theorem (1924), the radiative flux would temporarily
drop at this distortion, with an associated local drop in wind
momentum (for a radiatively-driven wind), yielding a longitudinal wind
asymmetry.

Another mechanism that could cause a longintudinally asymmetric wind
is proper motion of the mass-losing star through the ISM.  This was
first treated by \citet{Wea77} and more recently by e.g.\
\citet{Wil96} and \citet{Com98}.  A slow, westward motion of the
progenitor through the ISM would pile up more material at the western
wind-ISM contact discontinuity than to the east.  Similarly, the star
would be closer to this western discontinuity than to the west.  Such
an effect has been seen for PNe \citep{TK96} and for the LMC luminous
blue variable S119 \citep{DC01}.  A proper motion of less than 0.1
\kms is required to produce the observed offets, however, this
scenario is predicated on the progenitor not blowing an ISM bubble
during its MS that is larger than any of the CS structures revealed in
this paper.  We look forward to a new generation of hydrodynamical
models that will readdress the mechanisms of mass loss in Sk
$-69\degr$~202, and subsequently the phenomena of bipolar mass loss in
many more systems.

\smallskip

%%%%%%%%%%%%%%%%%%%%%%%%%%%%%%%%%%%%%%%%%%%%%%%%%%%%%%%%%%%%%%%%%%
% acknowledgement
%%%%%%%%%%%%%%%%%%%%%%%%%%%%%%%%%%%%%%%%%%%%%%%%%%%%%%%%%%%%%%%%%%

\acknowledgements

B.E.K.S. wishes to thank Alex Bergier, Eric Blackman, Roger Chevalier,
Adam Frank, Peter Lundqvist, Geralt Mellema, and Robert Uglesich.  We
gratefully acknowledge our referee, Richard McCray, for his critical
reading and insightful feedback on all the manuscripts in this
series. This research was based in part on observations made with the
NASA/ESA Hubble Space Telescope, obtained from the Data Archive at the
Space Telescope Science Institute, which is operated by the
Association of Universities for Research in Astronomy, Inc., under
NASA contract NAS 5-26555.  This work was generously supported by
STScI grants GO 8806, 8872, 9111, 9328, 9428, \& 9343; NASA
NAG5-13081; NSF AST 02 06048; and by Margaret Meixner and STScI DDRF
grant 82301.

%%%%%%%%%%%%%%%%%%%%%%%%%%%%%%%%%%%%%%%%%%%%%%%%%%%%%%%%%%%%%%%%%%
%                    bibliography                                %
%%%%%%%%%%%%%%%%%%%%%%%%%%%%%%%%%%%%%%%%%%%%%%%%%%%%%%%%%%%%%%%%%%

\end{document}